\documentclass[12pt]{iopart}
\usepackage[square,numbers]{natbib}
\usepackage{graphicx}	
\expandafter\let\csname equation*\endcsname\relax
\expandafter\let\csname endequation*\endcsname\relax
\usepackage{amsmath}
\usepackage{amssymb}	
\usepackage{xcolor}
\definecolor{navy}{RGB}{0,0,160}
\usepackage[colorlinks=true,allcolors=navy]{hyperref}
\usepackage{upgreek}
\usepackage{etoolbox} 
\makeatletter
\renewcommand\@appendixstar{\@@par
 \ifnumbysec 
 \@addtoreset{table}{section}
 \@addtoreset{figure}{section}\fi
 \setcounter{section}{0}
 \setcounter{subsection}{0}
 \setcounter{subsubsection}{0}
 \setcounter{equation}{0}
 \setcounter{figure}{0}
 \setcounter{table}{0}
 \def\thesection{\Alph{section}} 
 \def\theequation{\ifnumbysec
      \Alph{section}.\arabic{equation}\else
      \Alph{section}\arabic{equation}\fi}
 \def\thetable{\ifnumbysec
      \Alph{section}\arabic{table}\else
      A\arabic{table}\fi}
 \def\thefigure{\ifnumbysec
      \Alph{section}\arabic{figure}\else
      A\arabic{figure}\fi}}
\makeatother

\usepackage{mathptmx}

\usepackage{fancyhdr} 
\newcommand{\D}{\mathrm{D}}
\newcommand{\di}{\mathrm{d}}
\newcommand{\revision}[1]{{\bfseries #1}}
\renewcommand{\revision}[1]{#1}
\newcommand{\revisiontwo}[1]{{\color{red}#1}}
\renewcommand{\revisiontwo}[1]{#1}

\begin{document}

\title[Hamiltonians and canonical coordinates for spinning particles in curved space-time]{Hamiltonians and canonical coordinates for spinning particles in curved space-time}

\author{Vojt{\v e}ch Witzany,$^{1,2}$
Jan Steinhoff,$^{3}$
and Georgios Lukes-Gerakopoulos$^{1}$}
\address{$^{1}$Astronomical  Institute  of  the  Academy  of  Sciences  of  the  Czech  Republic,
Bo\v{c}n{\'i}  II  1401/1a,  CZ-141  00  Prague,  Czech  Republic}
\address{$^{2}$Center of Applied Space Technology and Microgravity (ZARM), 
Universit{\"a}t Bremen, 
Am Fallturm 2,  
D-28359 Bremen, Germany}
\address{$^{3}$Max Planck Institute for Gravitational Physics (Albert Einstein Institute), Am M{\"u}hlenberg 1, Potsdam D-14476, Germany}
\ead{vojtech.witzany@asu.cas.cz, jan.steinhoff@aei.mpg.de, gglukes@gmail.com}
\vspace{10pt}
\begin{indented}
\item[]\today
\end{indented}

\begin{abstract}
The spin-curvature coupling as captured by the so-called Mathisson-Papapetrou-Dixon (MPD) equations is the leading order effect of the finite size of a rapidly rotating compact astrophysical object moving in a curved background. It is also a next-to-leading order effect in the phase of gravitational waves emitted by extreme-mass-ratio inspirals (EMRIs), which are expected to become observable by the LISA space mission. Additionally, exploring the Hamiltonian formalism for spinning bodies is important for the construction of the so-called Effective-One-Body waveform models that should eventually cover all mass ratios.

The MPD equations require supplementary conditions determining the frame in which the moments of the body are computed. We review various choices of these supplementary spin conditions and their properties. Then, we give Hamiltonians either in proper-time or coordinate-time parametrization for the Tulczyjew-Dixon, Mathisson-Pirani, and Kyrian-Semer{\'a}k conditions. Finally, we also give canonical phase-space coordinates parametrizing the spin tensor. We demonstrate the usefulness of the canonical coordinates for symplectic integration by constructing Poincar{\'e} surfaces of section for spinning bodies moving in the equatorial plane in Schwarzschild space-time. We observe the motion to be essentially regular for EMRI-ranges of the spin, but for larger values the Poincar{\'e} surfaces of section exhibit the typical structure of a weakly chaotic system. A possible future application of the numerical integration method is the inclusion of spin effects in EMRIs at the precision requirements of LISA. 
\end{abstract}

%
%
%
%
%


\tableofcontents

\section{\label{sec:intro}Introduction}


The detection of black-hole and neutron-star binary inspirals by the aLIGO and aLIGO-Virgo detectors mark the dawn of gravitational-wave astronomy \citep{gw1,gwkatalog}. The equations of Einstein gravity are put to test not only by the phenomenon and detection of gravitational waves itself, but also by the precise shape of the detected signal \citep{gwtests}. Furthermore, the analysis of the signal from neutron-star binaries provides precious astrophysical information about their composition \citep{gwEOS1,gwEOS2,gwEOS3}, and the observations of the electromagnetic aftermath is key to the explanation of the origin of the energetically unfavorable heavy elements in our Universe \citep{kilonova,r-process}.

Upcoming space-based missions such as LISA promise to probe the gravitational-wave spectrum in lower frequencies than terrestrial detectors such as Advanced LIGO and Virgo and, thus, to explore the dynamics of many other types of sources of gravitational radiation \citep{amaro2017}. One such particular class of sources are the so-called extreme-mass-ratio inspirals (EMRIs), during which stellar-mass compact objects spiral into massive black holes, which have masses at least five orders of magnitude above the solar mass \citep{babak2017}.

Independent of the mass ratio between the components of the system, neither the primary nor the secondary of the binary can be modeled as point particles in an accurate treatment of the inspiral, and effects of the finite size of the bodies must be taken into account. This is clear in the case of binaries of comparable size and mass, but in the case of EMRIs a more careful argumentation must be given. 

Let us denote the mass of the primary massive black hole as $M$ and the mass of the secondary stellar-mass object as $\mu$. Then the mass-ratio in EMRIs is $q \equiv \mu/M \sim 10^{-4}-10^{-7}$ and one can describe the gravitational field of the secondary as a perturbation on top of the gravitational field of the primary. As a result, the secondary is usually described as moving on the original background while being subject to a self-force whose relative size with respect to the Christoffel-connection terms is of the order $\mathcal{O}(q)$ \citep[see][for reviews and a complete list of references]{poisson2011,barack2018}.

Now consider the effects of the finite size of the secondary. If the secondary is rotating at relativistic speeds, a matter element on its surface will feel a relative acceleration with respect to the center of mass that is proportional to the velocity of the surface $v$, the radius of the object $r$, and the local space-time curvature $R$. Under the assumption of a balance of forces inside the body, this will result in a ``spin force'' $\sim \mu v r R$ acting on the center of mass. Let us further assume that the binary orbital separation is within a few horizon radii of the primary, and that the secondary is either a maximally spinning black hole, a few-millisecond pulsar, or a few-second pulsar. We then get respectively  $vrR \sim 1 q/M, 10^{-1} q/M, 10^{-4} q/M$. When we consider that the Christoffel symbols scale as $\sim 1/M$, we see that the relative size of the acceleration caused by the spin force is then $\mathcal{O}(q)$, the same as the gravitational self-force.

The effects of the self-force and the spin force on the orbit will thus both scale as $\mathcal{O}(q)$ and would be essentially impossible to distinguish from a geodesic when using observables collected over just a few orbital periods. Nevertheless, the orbit will only decay over $\mathcal{O}(1/q)$ cycles and the small deviations amount to secular effects in the phase of the orbit. The final orbital phase $\phi_\mathrm{f}$ can then be schematically written as a sum of contributions of the form \citep{hinderer2008}
\begin{align}
\phi_{f} &= \phi^{(1)}_{\mathrm{avg}} && \mathcal{O}(q^{-1})
\\ & + \phi^{(1)}_{\mathrm{osc}} + \phi^{(2)}_{\mathrm{avg}} + \phi^{}_{\mathrm{spin}} && \mathcal{O}(1)
\\ & + \phi^{(2)}_{\mathrm{osc}} + \phi^{(3)}_{\mathrm{avg}} + \phi^{}_{\mathrm{quad}} && \mathcal{O}(q) \,
\\ & +... && \mathcal{O}(q^2) \,,
\end{align}
where ``avg'' and ``osc'' stand respectively for contributions from the averaged dissipative, and oscillating dissipative and conservative parts of the self-force computed from the metric perturbations of order $(n)$. Then, at the same order as the first-order conservative piece of the self-force appears the contribution of the spin force. Both the $\mathcal{O}(q^{-1})$ and the $\mathcal{O}(1)$ terms must be eventually included if sub-radian precision is to be achieved in the EMRI wave-form modeling. \revision{Concrete computations of $\phi_\mathrm{spin}$ were carried out by \citet{Warburton2017}, who showed that during the entire length of the inspiral the spin-force can lead to a dephasing as large as 15 cycles as compared to a waveform where the spin-force was neglected.}

The $\mathcal{O}(q)$ contributions to the phase then contain the contribution of the next-to-leading effect of the finite size of the secondary, the quadrupolar coupling. In particular, this will include the spin-induced quadrupole that scales as $\sim S^2$ for neutron stars and black holes \citep{hansen1974,laarakkers1999,steinhoff2015}, where $S \sim \mu r v$ is the spin magnitude. Tidal deformation of the body also formally appears in the quadrupole; however, it can be estimated to enter the equations of motion at relative order $\mathcal{O}(q^4)$ \citep{damour2009,binnington2009,steinhoff2012} and it will thus enter the phase only at $\mathcal{O}(q^3)$ for conservative effects and perhaps at $\mathcal{O}(q^2)$ if the dissipative tidal effects contribute to the orbital decay time.

In summary, we see that the spin-curvature coupling considered at least to linear order is an indispensable piece of any EMRI model. However, the spin-curvature coupling also plays an important role in the post-Newtonian (weak-field and slow-motion) description of comparable-mass binaries \cite{Blanchet2013, Schafer2018}; the conservative dynamics includes all fourth-order spin-induced effects so far \cite{Levi2016}. But still, a wave-form model that encompasses mass ratios from comparable to extreme is highly desirable. This is one goal of the effective-one-body (EOB) model \cite{Buonanno1998,Bohe2016,Nagar2018,Buonanno2014,Damour2016}, being probably the best candidate to succeed in this endeavor. While incorporating all $\mathcal{O}(1)$ self-force effects is progressing \cite{Akcay2012}, one flavor of EOB models already incorporates the test-spin force on a Kerr background to linear order in the test-spin via a Hamiltonian
\cite{barausse2009,Bohe2016},
the central piece encoding the conservative dynamics in any EOB model (but see the progress of the other EOB flavor in Refs. \citep{Damour2014,Harms2016,Kavanagh2017,Bini2018}). In this context, exploring simplified Hamiltonian descriptions of spinning bodies appears to be crucial. 

In this paper, we study the Hamiltonian formalism for a spinning particle moving in a given space-time metric.
\revisiontwo{The so-called Mathisson-Papapetrou-Dixon (MPD) equations, which capture the effects of the spin-force on the orbit, clearly form a conservative system. However, their relation to the Hamiltonian formalism was previously established only partially. The MPD equations require a condition that specifies the referential world-line inside the body, the so-called spin supplementary condition. We complete the previous works \cite{barausse2009,vines2016} by providing Hamiltonians for all commonly used covariant or ``comoving'' supplementary conditions. Additionally, we map the corresponding phase space with canonical coordinates, which allows for efficient numerical integration. An important part of our paper is a discussion of the physical and redundant ``spin-gauge'' degrees of freedom that appear in the equations. In particular, we obtain simple and elegant expressions at the cost of a spin-gauge degree of freedom remaining in our phase space.
}

In Section \ref{sec:mpd} we review the MPD equations and their properties under various supplementary spin conditions.
We then proceed to the Hamiltonian formalism in Section \ref{sec:ham}. We present the Poisson brackets and various sets of variables that can be used during the evolution, and Hamiltonians for all the usual comoving supplementary conditions both in proper-time and coordinate-time parametrizations. Next, in Section \ref{sec:canon}, we also give a set of canonical coordinates covering the spin tensor. Finally, in Section \ref{sec:plan}, we demonstrate the power of the new coordinates and Hamiltonian formalism by numerically studying spinning particles moving in the equatorial plane of a Schwarzschild black hole. The paper also contains a number of Appendices that provide context to the presented results and details of the derivations mentioned in the main text.

We use the $G=c=1$ geometrized units and the (-+++) signature of the metric. Our convention for the Riemann tensor $R^{\mu}_{\;\nu\alpha\beta}$ is such that $2 a_{\mu;[\alpha\beta]} = R^{\nu}_{\;\mu\alpha\beta} a_{\nu}$ for a generic $a_\mu$, or explicitly $R^{\mu}_{\;\nu\alpha\beta} = 2 \Gamma^{\mu}_{\;\rho [\alpha} \Gamma^{\rho}_{\;\beta] \nu} -2 \Gamma^{\mu}_{\;\nu [\alpha , \beta]}$. The anti-symmetrization of a tensor is written as $W_{[\alpha\beta]}=\frac{1}{2}\left(W_{\alpha\beta} -W_{\beta\alpha} \right)$, while the symmetrization as $W_{(\alpha\beta)}=\frac{1}{2}\left(W_{\alpha\beta} +W_{\beta\alpha} \right)$. We denote the covariant time derivative by an overdot, $\dot{A}^{\mu \nu...}_{\gamma \delta ...} \equiv \mathrm{D} A^{\mu \nu ...}_{\gamma \delta ...}/ \di \tau \equiv  A^{\mu \nu ...}_{\gamma  \delta ...;\kappa} \dot{x}^\kappa$. $\eta^{\mu\nu}$ with any indices is the Minkowski tensor, and ${\delta^\mu}_\nu$ denotes the Kronecker delta.


\section{MPD equations} \label{sec:mpd}

The equations of motion of massive bodies in a gravitational field is among the most basic topics in Newtonian mechanics, and among the toughest problems in general relativity. Surprisingly, just assuming the covariant conservation of energy-momentum of the body restricts these equations to be of the celebrated MPD form in general relativity. The MPD equations to pole-dipole order read \citep{mathisson1937,papapetrou1951,dixon1964}
\begin{subequations}
\label{eq:mpd}
\begin{align}
&\dot{P}^\mu = -\frac{1}{2} R^{\mu}_{\; \nu \kappa \lambda} \dot{x}^\nu S^{\kappa \lambda} \,,\\
&\dot{S}^{\kappa \lambda} = P^\kappa \dot{x}^\lambda - P^\lambda \dot{x}^\kappa\,,
\end{align}
\end{subequations}
where $x^\mu(\tau)$ is the world-line of some representative centroid from within the rotating body, $S^{\kappa \lambda}$ the spin tensor, and $P^\mu$ the momentum (flux of stress-energy) of the body. 
Here $\tau$ is the proper time, $\dot x^\mu \dot x_\mu = -1$.
But it is noteworthy that the MPD equations are invariant under affine reparametrizations of the world-line. 

The relation between $\dot{x}^\nu$ and $P^\nu$ is underdetermined and has to be derived from a supplementary spin  condition. A supplementary spin condition is usually given in the form $S^{\mu\nu}V_\nu=0$, where $V_\nu$ is some time-like vector. The physical interpretation of this supplementary condition is that $V^\nu$ is the frame in which the momenta of the stress-energy tensor $P^\mu$ and $S^{\nu \kappa}$ are computed, and the position of the referential world-line $x^\mu(\tau)$ is then the center of mass of the spinning body in this frame \citep{costa2015}. 

The MPD equations as stated here do not include the contributions from the quadrupole and higher-order mass moments of the body. They are in fact universal at pole-dipole order, i.e., independent of the internal structure of the body. Amongst other effects and as already mentioned in the Introduction, one expects the rotation to deform the body and thus produce a structure-dependent quadrupole moment that scales as $S^2$; this holds in particular for rotating black holes and neutron stars. Since we are not including such spin-quadratic terms in the equations either way, it is often meaningful to truncate the formulas at some low order in $S$.

Some of the identities that are useful independent of the supplementary condition read
\begin{align}
& \dot{x}^{(\mu}\dot{S}^{\nu \kappa)_\mathrm{cycl.}} = 0 \,  \label{eq:scykl},\\
& P^\mu = m \dot{x}^\mu + \dot{x}_\gamma \dot{S}^{\gamma \mu}\,, \label{eq:pxdot}\\ 
& m \equiv -P_\mu \dot{x}^\mu\,. \label{eq:mdef}
\end{align}
A number of other useful identities along with a brief historical review of the MPD equations can be found in Ref. \citep{semerak1999}.

We also define the spin vector $s^\mu$, the spin magnitudes $S, S^*$, and a mass-like quantity $\mathcal{M}$ by
\begin{align}
& s^\mu \equiv -\frac{1}{2 \sqrt{-V^\alpha V_\alpha}} \epsilon^{\mu \nu \kappa \lambda} V_\nu S_{\kappa \lambda} = -\frac{1}{\sqrt{-V^\alpha V_\alpha}}{\star S}^{\mu\nu} V_\nu\,,\\
& S\equiv \sqrt{\frac{1}{2} S^{\kappa \lambda} S_{\kappa \lambda}} = s^\mu s_\mu\,, \\
& S^*\equiv \sqrt{{\star S}^{\kappa \lambda} S_{\kappa \lambda}} = \sqrt{\frac{1}{2} S^{\mu\nu} S^{\kappa \lambda} \epsilon_{\mu\nu\kappa\lambda}} \,, \\
& \mathcal{M} \equiv \sqrt{- P^\alpha P_\alpha} \,,
\end{align}
where ${\star S}^{\mu \nu} \equiv \epsilon^{\mu\nu \kappa \lambda} S_{\kappa \lambda}/2 = (s^\mu V^\nu - s^\nu V^\mu)/\sqrt{-V^\alpha V_\alpha}$.
It should be noted that the definition of $s^\mu$ will be different whenever a different supplementary condition is chosen.

Now we see that $S^{\kappa \lambda}s_\lambda = 0$ and we can build a projector on the sub-space orthogonal to $V_\mu, s_\nu$ as
\begin{align}
h^\mu_{\;\nu} = \frac{1}{S^2} S^{\mu\kappa} S_{\nu \kappa} = \left(\delta^\mu_{\;\nu} + \frac{V^\mu V_\nu}{(-V^\alpha V_\alpha)} - \frac{s^\mu s_\nu}{S^2} \right)\,. \label{eq:proj}
\end{align}
\revision{Let us now briefly discuss the independent components of the spin tensor. A general anti-symmetric tensor in 4 space-time dimensions will have 6 independent components. However, the fact that there exists a $V^\mu$ such that $S^{\mu\nu}V_\nu = 0$ implies $S^* =0$ and this reduces the number of independent components to 5. 

In the cases where the spin magnitude $S$ is a constant of motion, the number of independent components of $S^{\mu\nu}$  drops to 4. One such particular parametrization is given in Section \ref{sec:canon}.

However, the time derivative of $S$ is given by
\begin{align}
\frac{\mathrm{d}}{\di \tau} (S^2) = S_{\mu\nu}\dot{S}^{\mu\nu}=2 S_{\mu\nu} P^\mu  \dot{x}^\nu\,. \label{eq:smagev}
\end{align}
The expression \eqref{eq:smagev} might be zero or non-zero, depending on the supplementary condition, and the choice of the most practical parametrization of the spin tensor will thus vary accordingly.  
}

Now, the question is which supplementary spin condition should be adopted to close the system of MPD equations. The best answer that one can give, however, is that virtually any condition is physically viable, at least at the (universal) pole-dipole order. Hence, in the remaining part of this section, we review all commonly proposed classes of supplementary spin conditions, \revision{focusing in particular on the phase space of the resulting evolution equations. Specifically, by ``phase space'' and ``phase-space dimension'' we will mean the number of variables that need to be stored and updated during every step of a numerical integration of the equations.

The fundamental issue with the MPD system of equations is the fact that there is no ``retrospective'' way to directly verify the physicality of their approximation once we have replaced the full, original continuum of the body with the multipole momenta $P^\mu$ and $S^{\mu\kappa}$, and the representative world-line $x^\mu(\tau)$. Furthermore, there is no supplementary condition that can be deemed as universally physical in any situation. Thus, we must judge the physicality of the supplementrary spin conditions on a case-to-case basis, using mainly criteria of self-consistency.} 


\subsection{The KS condition} \label{subsec:ks}

Eq.~\eqref{eq:pxdot} indicates that the momentum is generally linearly independent of the four-velocity.
\citet{kyrsem} (KS) asked the question under which supplementary condition is the momentum proportional to the four-velocity, $P^\mu = m \dot{x}^\mu$, and found that this is true when we assume the existence of a time-like vector $w_\mu$ such that $S^{\mu \nu} w_\nu = 0$ and $\dot{w}_\nu = 0$. We also conventionally set $w_\alpha w^\alpha = -1$. The MPD equations then simplify into the form 
\begin{subequations}
\label{eq:ks}
\begin{align}
&P^\mu = m \dot{x}^\mu\,, \label{eq:kspu}\\
&\dot{m} = 0\,, \label{eq:ksm}\\
&\ddot{x}^\mu = - \frac{1}{2m} R^\mu_{\;\nu \kappa \lambda} \dot{x}^\nu S^{\kappa \lambda}   \,, \label{eq:ksp}\\
&\dot{S}^{\kappa \lambda} = 0 \,. \label{eq:ksS}
\end{align}
\end{subequations}
This is probably the simplest form of the MPD equations one can acquire and it can in fact be generated by a large set of other supplementary conditions, which is discussed in Appendix \ref{app:ks}. 

\revision{The physical meaning of the vector $w^\mu$ is clear. It corresponds to a situation where one first chooses an arbitrary frame in which the moments of the body are computed, and then transports this frame along the world-line for further moment computations. However, the arbitrariness of this initial choice can be problematic because of two reasons. 

First, there are then many formal initial conditions that correspond to the same physical evolution. (How such an equivalence is established has been discussed in Refs. \citep{kyrsem,vines2016,costa2017}.) Second, the frame $w^\mu$ can be arbitrarily boosted with respect to the real motion of the bulk of the matter. In return, the respective center of mass can then be even outside the body and the multipolar MPD approximation is broken \citep{moller1949}. A consistency criterion for the physicality of a KS evolution thus is that $\dot{x}^\mu w_\mu \sim -1$ so that the boost of the averaging frame with respect to the body frame is only moderate. 

Another criterion of almost equivalent meaning that, nonetheless, does not refer to $w^\mu$ is $|S^{\mu\nu}\dot{x}_\nu/S| \lesssim 1$. Let us define for every given value of the time parameter $\tau$ a local fiducial choice of the frame that has exactly $w^\mu_\mathrm{f}(\tau) = \dot{x}^\mu(\tau)$. Then the criterion $|S^{\mu\nu}\dot{x}_\nu/S| \lesssim 1$ essentially states that the shift of our actual centroid $x^\mu(\tau)$ from the fiducial centroid $x^\mu_\mathrm{f}(\tau)$ has to be less than the so-called M\"{o}ller radius \citep{moller1949}.}

Now, the system of equations \eqref{eq:ks} is characterized by a phase space $(x^\mu, P^\nu, S^{\kappa \lambda})$. An important point is to realize that once an initial condition with some vanishing direction of the spin tensor is chosen, $S^{\mu\nu} w_\nu|_{\tau = \tau_0} = 0$, the equations of motion \eqref{eq:ks} will evolve with two vanishing directions (the first one proportional to $w^\mu$, and the second one proportional to $s^\mu$) in a way so that we can always choose for one of them to fulfill $\dot{w}^\mu = 0$. 

In other words, once the initial condition is set up with a degenerate spin tensor, the set of equations \eqref{eq:ks} can be evolved at face value without further reference to the auxiliary vector $w^\mu$. \revision{By comparing with equation \eqref{eq:smagev} we see that the spin magnitude $S$ is conserved, $\dot{S} = 0$, and the dynamical phase space of the problem thus has 4 new dimensions in the spin sector on top of the usual orbital degrees of freedom.}

The equations of motion can also be re-expressed using $w_\nu$ and the respective spin vector $s_\mu$ as 
\begin{subequations}
\label{eq:ksws}
\begin{align}
&\ddot{x}^\mu =  \frac{1}{m} 
{\star R}^\mu_{\;\nu \kappa \lambda} \dot{x}^ \nu s^\kappa w^\lambda\,, \\
&\dot{w}^\kappa = \dot{s}^\lambda = 0 \,, 
\end{align}
\end{subequations}
where ${\star R}_{\mu\nu \kappa \lambda} \equiv R_{\mu\nu \gamma \delta} \epsilon^{\gamma \delta}_{\;\;\; \kappa \lambda}/2$.
In this case the phase space $(x^\mu, \dot{x}^\nu,s^\lambda, w^\kappa)$ consists of the coordinate positions, velocities, the spin vector, and the auxiliary vector $w^\lambda$.


\subsection{The MP condition} \label{subsec:mp}

Another supplementary spin condition considered by various authors \citep{frenkel1926,mathisson1937,pirani1956} is $S^{\mu\nu} \dot{x}_\nu = 0$. We will call it the Mathisson-Pirani (MP) spin condition due to the pioneering works using this condition in the context of curved space-time \citep{mathisson1937,pirani1956}; in the context of flat space-time, it is often called the Frenkel spin condition due to the pioneering work of \citet{frenkel1926}. Under this supplementary condition, the MPD equations are simply the equations \eqref{eq:mpd} with the substitution of the following relation in place of $\dot{x}^\mu$ \citep{costa2017}
\begin{align}
\dot{x}^\mu = \frac{1}{m} P^\nu\left(\delta^\mu_\nu - \frac{1}{S^2} S^{\mu\kappa}S_{\nu \kappa}\right) = P^\nu\left(\delta^\mu_\nu - h^\mu_{\;\nu}\right) \,. \label{eq:mppu}
\end{align}
Once again, in this representation the phase space needed for numerical evolution is $(x^\mu,P^\nu,S^{\kappa\lambda})$ \revision{and we see that the supplementary condition along with \eqref{eq:smagev} implies $\dot{S} = 0$. Additionally, we notice from \eqref{eq:mppu} that the expression $S^{\mu\nu}\dot{x}_\nu = 0$ will be fulfilled by any choice of initial data $(x^\mu,P^\nu,S^{\kappa\lambda})$ as long as $S^{\mu\nu}$ is degenerate in {\em some} time-like direction. We thus again conclude that the spin sector of the phase space adds 4 new dimensions on top of the usual orbital degrees of freedom.}

Another representation of the phase space is through the spin vector and higher order derivatives of the position:
\begin{align}
&\dddot{x}^\mu = f^\mu(x^\nu, \dot{x}^\lambda,\ddot{x}^\kappa,s^\gamma)\,,\\
&\dot{s}^\lambda = s^\nu \ddot{x}_\nu \dot{x}^\lambda\,, 
\end{align}
where $f^\mu$ is derived in Appendix \ref{app:mp} and its explicit form is given in equation \eqref{eq:trojak}. In other words, the phase space in this description consists of $(x^\mu,\dot{x}^\nu,\ddot{x}^\kappa,s^\lambda)$. When we compare these variables with that of the KS condition, we see that even though we are not evolving any auxiliary $w^\lambda$, we do, however, store additional data in the acceleration vector $\ddot{x}^\lambda$. 

\revision{The relation $S^{\mu\nu}\dot{x}_\nu = 0$ or $V^\mu \propto \dot{x}^\mu$ can be read as an implicit condition on the frame $V^\mu$ in which the momenta and center of mass are computed: For every value of the time parameter $\tau$, it is to be the frame in which the resulting center of mass $x^\mu(\tau)$ is at rest. Unfortunately, this is not sufficient to determine the frame $V^\mu$ uniquely! In return, there will be many initial conditions under the MP+MPD system that will describe the same physical evolutions in very much the same way as in the case of the KS condition. A recent detailed discussion of this degeneracy was given by \citet{costa2017}. Using their examples, it would seem that a good consistency criterion for the MP+MPD system is that the stress-energy fluxes are not too misaligned with the center-of-mass motion, $\dot{x}^\mu P_\mu/\mathcal{M} \sim -1$.}


\subsection{The TD condition} \label{subsec:td}
The Tulczyjew-Dixon (TD) supplementary spin condition \citep{tulczyjew1959,dixon1970} $S^{\mu\nu} P_\nu = 0$ leads to the MPD equations of motion where we substitute $\dot{x}^\mu$ throughout by \citep{ehlers1977,obukhov2011}
\begin{align}
& \dot{x}^\mu = \frac{m}{\mathcal{M}^2} \left( P^\mu + \frac{2 S^{\mu\nu} R_{\nu \gamma \kappa \lambda} P^\gamma S^{\kappa \lambda} }{4 \mathcal{M}^2 +  R_{\chi \eta \omega \xi} S^{\chi \eta} S^{\omega \xi}} \right)\,, \label{eq:tdup}
\end{align}
where $\mathcal{M}$ is an integral of motion, $\dot{\mathcal{M}} = 0$. The other mass $m$ is not an integral of motion, and can be easily expressed as a function of $P^\mu, S^{\kappa \lambda}, R_{\alpha \beta \gamma \delta}$ from $\dot{x}^\mu \dot{x}_\mu = -1$ as
\begin{align}
&m= \frac{\mathcal{A} \mathcal{M}^2}{\sqrt{ \mathcal{A}^2 \mathcal{M}^2 - \mathcal{B} S^2}} \,, \label{eq:mtd}\\
&\mathcal{A} = 4 \mathcal{M}^2 + R_{\alpha \beta \gamma \delta} S^{\alpha \beta} S^{\gamma \delta} \,, \\
&\mathcal{B} = 4 h^{\kappa \eta} R_{\kappa \iota \lambda \mu } P^\iota S^{\lambda \mu} R_{\eta \nu \omega \pi } P^\nu S^{\omega \pi} \,.
\end{align}
The phase space is then parametrized by $(x^\mu,P^\nu,S^{\kappa\lambda})$, we again obtain from \eqref{eq:smagev} that $\dot{S} = 0$, \revision{and the heuristic conditions on the physical validity of the MPD+TD system can be cast as $-\dot{x}^\mu P_\mu/\sqrt{-P^\alpha P_\alpha} = m/\mathcal{M} \sim 1$, or at least that $m$ is real \citep{semerak1999}}. 

\revision{However, in this case we notice that the spin tensor and the momentum are not linearly independent and the number of new independent phase space dimensions in the spin sector will be smaller than 4. The relation $S^{\mu\nu}P_\nu = 0$ has 4 components out of which one is trivial, $S^{\mu\nu}P_\mu P_\nu = 0$ and one is equivalent to $S^* = 0$. Thus, by counting the independent dynamical components of the spin tensor, we see that the TD+MPD system adds only 2 active phase space dimensions on top of the orbital degrees of freedom.}

Once again, there is the possibility to transform to a spin vector which yields \citep[cf.][]{suzuki1997}
\begin{align}
&\dot{P}^\mu =  \frac{1}{\mathcal{M}} {\star R}^\mu_{\;\nu\kappa \lambda} \dot{x}^\nu s^\kappa P^\lambda \,,\\
&\dot{s}^\mu = \frac{1}{ \mathcal{M}^3} {\star R}_{\gamma \nu \kappa \lambda} s^\gamma \dot{x}^\nu s^\kappa P^\lambda P^\mu\,,
\end{align} 
where we use equation \eqref{eq:tdup} to eliminate $\dot{x}^\nu$. This set of equations is non-linear and complicated, but the phase space is now composed only of $(x^\mu, P^\nu, s^\kappa)$, which is probably the most economic covariant formulation of the equations of motion.


\subsection{The CP and NW conditions} \label{subsec:cpnw}

The Corinaldesi-Papapetrou (CP) \citep{corinaldesi2003} and Newton-Wigner (NW) \citep{pryce1948,newton1949} conditions employ an external time-like vector field $\xi^\mu(x^\nu)$ in the supplementary condition
\begin{align}
S^{\mu\nu}\left(\xi_\nu + \alpha \frac{P_\nu}{\mathcal{M}} \right) = 0\,,
\end{align}
where $\alpha = 0$ corresponds to the CP and $\alpha=1$ to the NW condition. \revision{The physical interpretation of the CP condition is simply that the averaging frame is some ``lab frame'' $\xi^\mu$ that is, unlike in the case of the KS condition, defined by a background structure rather than local transport with the body. On the other hand, the Newton-Wigner condition does not have any simple physical interpretation. Similarly to the KS condition, one is entirely free to choose $\xi^\nu$, and these choices are physically valid only if the particle is non-relativistic with respect to the frame $(\xi^\nu + \alpha P^\nu/\mathcal{M})$, $\dot{x}^\mu(\xi_\mu + \alpha P_\mu/\mathcal{M}) \sim -1$.}

The convenience of these supplementary conditions lies in the fact that one can recast the evolution for the spin tensor in terms of a tetrad basis $S^{AB} = e_\mu^A e_\nu^B S^{\mu\nu}$ and by choosing for instance $e^0_\mu = \xi_\mu$ we can eliminate 3 of the six independent spin-tensor components $S^{0I}, I=1,2,3$ as
\begin{equation}
S^{0I} = -\frac{\alpha}{\mathcal{M} + \alpha P_0} P_J S^{JI}\,,\; J=1,2,3 \,. \label{eq:nwcons}
\end{equation}
The equations of motion for the spin tensor are obtained with the help of \eqref{eq:scykl} as
\begin{align}
\dot{S}^{\mu\nu} = 2 S^{\kappa [\mu}\dot{x}^{\nu]} \frac{(\mathcal{M}\xi_{\kappa;\lambda} - \alpha R_{\hat{\kappa}\lambda\gamma \delta}S^{\gamma \delta}/2)\dot{x}^\lambda}{(\mathcal{M}\xi_\chi + \alpha P_\chi)\dot{x}^\chi}\,,
\end{align}
where the notation $\hat{\kappa}$ in the curvature tensor signifies the part orthogonal to $P^\nu$.

The momentum-velocity relation then attains the following implicit form
\begin{align}
m \dot{x}^\mu = P^\mu - (S^{\kappa \mu} + S^{\kappa \omega}\dot{x}_\omega \dot{x}^\mu) \frac{(\mathcal{M}\xi_{\kappa;\lambda} - \alpha R_{\hat{\kappa}\lambda\gamma \delta}S^{\gamma \delta}/2)\dot{x}^\lambda}{(\mathcal{M}\xi_\chi + \alpha P_\chi)\dot{x}^\chi}\,.
\end{align}
This relation is not exactly reversible into a $\dot{x}^\mu(P_\nu)$ or $P_\nu(\dot{x}^\mu)$ formula in the general case and one thus cannot always use the CP/NW condition to give a set of evolution equations in strictly closed form. 

Nevertheless, it is possible to iterate the momentum-velocity relation by starting from $\dot{x}^\mu = P^\mu/m + \mathcal{O}(S)$ to obtain results of higher and higher precision with respect to  powers of $S$. The first iteration yields
\begin{align}
m \dot{x}^\mu = P^\mu - \left( S^{\kappa \mu} + \frac{1}{m^2} S^{\kappa \omega}P_\omega P^\mu \right) \frac{\xi_{\kappa;\lambda}P^\lambda}{(m \xi_\chi + \alpha P_\chi)P^\chi} \label{eq:nwlin} + \mathcal{O}(S^2)\,. 
\end{align}

\revision{Finally, one can also eliminate the number of variables by considering that a ``projected spin magnitude'' is approximately conserved under the NW/CP evolution:}
\begin{align}
\frac{\di}{\di \tau} \left( \frac{1}{2}S^{\mu\nu}S_{\mu\nu} - S^{\kappa\gamma}S_{\gamma \delta} \dot{x}^\delta \dot{x}_\kappa \right) = 0 + \mathcal{O}(S^3)\,. \label{eq:smod}
\end{align}
Formulas such as \eqref{eq:nwlin} inserted into the MPD equations along with the assumption that \eqref{eq:nwcons} is exactly true at all times lead to closed-form evolution equations with the phase space $(x^\mu,P_\nu, S^{IJ})$. \revision{Then, by considering the approximately conserved projected spin magnitude \eqref{eq:smod}, we see that the total number of active phase-space dimensions in the spin sector is 2, similar to the TD condition.\footnote{On the other hand, one can notice from the residual terms in eq. \eqref{eq:smod} that by reducing with respect to the projected spin magnitude leads to a discarding of terms already of order $\mathcal{O}(S^{3/2})$ in the equations of motion rather than $\mathcal{O}(S^{2})$.}}

One other reason the NW condition in particular received heightened attention in the recent years is the fact that it can be formulated as a Hamiltonian system with the canonical $\mathrm{SO}(3)$ commutation relations for the spin vector \citep{hanson1974,barausse2009,vines2016}.



\section{Hamiltonians for spinning particles} \label{sec:ham}

In this section we construct Hamiltonian formulations of the MPD equations supplemented by various choices of spin conditions. Besides being of fundamental interest, these are often advantageous for certain applications. For instance, in forthcoming sections we study a numerical integration of the MPD equations using efficient symplectic integrators on phase space. EOB waveform models use Hamiltonians to encode the conservative binary dynamics since they can naturally be mapped between the case of two bodies and a reduced mass in a fixed (effective) background.


\subsection{The Poisson brackets} \label{subsec:poiss}

Before we are able to discuss Hamiltonians, we need to set up the stage in the form of a phase space
endowed with a Poisson bracket.
Consider the set of non-zero Poisson brackets for the phase-space coordinates $x^\mu, P_\nu, S^{\gamma \kappa}$
\begin{subequations}
\label{eq:poiss}
\begin{align}
\{x^\mu,x^\nu\} &= 0\,,\\
 \{x^\mu, P_\nu\} &=  \delta^\mu_\nu \label{eq:poissxpcov} \,, \\
\{P_\mu, P_\nu\} & = -\frac{1}{2} R_{\mu\nu \kappa \lambda} S^{\kappa \lambda} \label{eq:poisspp} \,, \\
 \{S^{\mu\nu}, P_\kappa\} & = -\Gamma^\mu_{\;\lambda \kappa} S^{\lambda \nu} - \Gamma^\nu_{\; \lambda \kappa} S^{\mu \lambda} \label{eq:poissSp} \,, \\
 \{S^{\mu\nu}, x^\kappa\} & = 0 \label{eq:poissSx} \,, \\
\begin{split}
 \{S^{\mu\nu}, S^{\kappa \lambda}\} & =  g^{\mu \kappa}S^{\nu \lambda} - g^{\mu \lambda}S^{\nu \kappa} + g^{\nu \lambda}S^{\mu \kappa} \\ & - g^{\nu \kappa}S^{\mu \lambda} \label{eq:poissSS} \,.
\end{split}
\end{align}
\end{subequations}
This set of brackets \revision{can be easily shown to fulfill Jacobi identities and thus to define a valid Poisson structure}. The brackets arise in many models for spinning-particle dynamics \citep{kunzle1972,feldman1980,tauber1988,kriplovich1989,barausse2009,ramirez2015,dambrosi2015}. Furthermore, it is easy to prove that the Poisson brackets follow from the generic effective action used in Refs. \citep{steinhoff2009,steinhoff2015,vines2016} (see Appendix \ref{app:canon}).

The Poisson brackets \eqref{eq:poiss} can be partially canonicalized by choosing an orthonormal tetrad $e_\mu^A\,,\; e_\mu^A e^{\mu B} = \eta^{AB}$ ($=$ Minkowski metric), and adopting a set of variables \citep{feldman1980,tauber1988,kriplovich1989}
\begin{align}
& S^{AB} = S^{\mu\nu} e_\mu^A e_\nu^B\,, \label{eq:stetr}\\ 
& p_\mu = P_\mu + \frac{1}{2}e_{\nu A;\mu}e^\nu_B S^{AB} \,. \label{eq:pnonc}
\end{align}
Under this change of variables the only non-zero brackets read
\begin{subequations}
\label{eq:partcan}
\begin{align}
 \{x^\mu, p_\nu\} &=  \delta^\mu_\nu \label{eq:poissxp} \,, \\
 \begin{split}
 \{S^{AB},S^{CD}\} &= \eta^{AC} S^{BD} - \eta^{AD} S^{BC} + \eta^{BD} S^{AC} - \eta^{BC} S^{AD} \,.
 \end{split}
\end{align}
\end{subequations}
In this coordinate basis it is clear that $S^{AB}$ and its commutation relations are a representation of the generators of the Lorentz group. Additionally, we see that the spin magnitudes $ S^2 = S^{AB}S_{AB}/2$ and $(S^*)^2 = S^{AB} S^{CD} \epsilon_{ABCD}$ are Casimir elements of this algebra. That is, the spin magnitudes $S, S^*$ commute with all the phase-space coordinates and will always be integrals of motion independent of the Hamiltonian.

\revision{
If we compare this with the fact that $\dot{S} \neq 0$ for the NW/CP condition, we see that this set of Poisson brackets {\em cannot} be used to evolve the MPD+NW/CP system of equations. We will thus only consider the KS, MP, and TD conditions in the rest of this Section. 

To prevent any confusion, we would like to stress that all equations of motion in this paper are computed exclusively with the brackets \eqref{eq:poiss} and their transformations such as \eqref{eq:partcan}. Specifically, we never use or explicitly derive a constrained (Dirac-Poisson) bracket, even though we discuss this prospect in Subsection \ref{subsec:whycan}.}


\subsection{Hamilton's equations of motion} \label{subsec:hameq}

We are now in a position to study the equations of motion for a generic Hamiltonian $H(x^\mu, P_\nu, S^{\kappa \lambda})$ with the Poisson brackets \eqref{eq:poiss}. We obtain
\begin{subequations}
\label{eq:ham}
\begin{align}
& \frac{\di x^\mu}{\di \lambda} = \frac{\partial H}{ \partial P_\mu} \,, \label{eq:hamvel}\\
\begin{split}
&\frac{\di P_\nu}{\di \lambda} + \frac{\partial H}{\partial x^\nu} - \frac{\partial H}{ \partial S^{\mu \kappa}} (\Gamma^\mu_{\;\nu\gamma} S^{\gamma \kappa} + \Gamma^\kappa_{\; \nu \gamma} S^{\mu \gamma}) =  - \frac{1}{2} R_{\nu \omega \lambda \chi} \frac{\partial H}{\partial P_\omega} S^{\lambda \chi}\,,
\end{split}\\
\begin{split}
&\frac{\di S^{\gamma \kappa}}{\di \lambda} + \Gamma^\gamma_{\;\nu \lambda} \frac{\partial H}{\partial P_\nu} S^{\lambda \kappa}+ \Gamma^\kappa_{\;\nu \lambda} \frac{\partial H}{\partial P_\nu} S^{\gamma\lambda} = \frac{\partial H}{\partial S^{\mu \nu}} (g^{\gamma \mu} S^{\kappa \nu} - g^{\gamma \nu} S^{\kappa \mu} + g^{\kappa \nu} S^{\gamma \mu} - g^{\kappa \mu} S^{\gamma \nu}) \,,
\end{split}
\end{align}
\end{subequations}
where $\lambda$ is some parameter along the trajectory.
These equations cannot be expected to make any sense on the full phase space, but only on the part where some supplementary condition $S^{\mu\nu} V_\nu = 0$ holds.

By comparison with equations \eqref{eq:mpd}, the equations \eqref{eq:ham} will be the MPD equations when the following equalities are fulfilled
\begin{align}
& \frac{\partial H}{\partial S^{\mu \nu}} (g^{\gamma \mu} S^{\kappa \nu} +\mathrm{perm.}) \cong P^\kappa \frac{\partial H}{\partial P_\gamma} - P^\gamma \frac{\partial H}{\partial P_\kappa} \,,\\
& \frac{\partial H}{\partial x^\nu} - \frac{\partial H}{ \partial S^{\mu \kappa}} (\Gamma^\mu_{\;\nu\gamma} S^{\gamma \kappa} + \Gamma^\kappa_{\; \nu \gamma} S^{\mu \gamma}) \cong - \Gamma^\alpha_{\;\beta \nu} \frac{\partial H}{\partial P_\beta} P_\alpha \,,
\end{align}
where $\cong$ means that the equalities need to hold only on a certain ``on-shell'' part of the phase space where conditions such as $S^{\mu\nu} V_\nu = 0$ hold. \revision{Naturally, this also means that we need to require that the supplementary condition itself is conserved, or}
\begin{align}
\{S^{\mu\nu}V_\nu(x^\kappa,P_\lambda,S^{\omega \pi}),H\} \cong 0\,.
\end{align}
The fact that the equalities are $\cong$ makes them impractical to solve directly and we resort to heuristic approaches.


\subsection{Hamiltonian for KS condition} \label{subsec:hamks}
\citet{kriplovich1989} postulated the following Hamiltonian for semi-classical spinning particles which is to be used along the Poisson brackets \eqref{eq:poiss} (see also \citet{dambrosi2015})
\begin{equation}
H_\mathrm{KS} = \frac{1}{2m} g^{\mu\nu} P_\mu P_\nu\cong-\frac{m}{2} \,. \label{eq:kripham}
\end{equation}
However, at the time of the publication of this Hamiltonian it was not clear what is the relation of the generated set of equations with the MPD equations. Nevertheless, we can now compare the generated equations of motion \eqref{eq:ham} with those corresponding to the relatively recently discovered KS supplementary spin condition \eqref{eq:ks} to see that the two sets of equation agree. 

In other words, the Hamiltonian \eqref{eq:kripham} generates the MPD equations under the KS spin condition. \revision{As already discussed in Subsection \ref{subsec:ks}}, the only requirement that needs to be fulfilled by the initial condition apart from four-velocity normalization is for $S^{\mu\nu}$ to have some vanishing time-like direction $w^\nu$, $S^{\mu\nu} w_\nu =0$.


\subsection{Hamiltonian for TD condition} \label{subsec:hamtd}
Our initial heuristic is to simply reproduce the momentum-velocity relation under the TD condition and see whether this is sufficient to determine the correct Hamiltonian. We take the velocity-momentum relation \eqref{eq:tdup} and combine it with \eqref{eq:hamvel} to obtain
\begin{align}
\frac{\partial H}{\partial P_\nu} \cong \frac{m}{\mathcal{M}^2} \left( P^\nu + \frac{2 S^{\nu\mu} R_{\mu \gamma \kappa \lambda} P^\gamma S^{\kappa \lambda} }{4 \mathcal{M}^2 +  R_{\chi \eta \omega \xi} S^{\chi \eta} S^{\omega \xi}}  \right) \label{eq:hamveltd} \,.
\end{align}
Now let us assume that the equations of motion hold under the on-shell conditions $\mathcal{M} = \sqrt{-P^\alpha P_\alpha}\,,\; S^{\mu\nu} P_\nu = 0$, where $\mathcal{M}$ is now some chosen constant independent of phase-space coordinates. Then the following holds
\begin{align}
 & \frac{\partial \,}{\partial P_\omega} \left[ (g^{\mu \nu} P_\mu P_\nu + \mathcal{M}^2) F \right] \cong 2F P^\omega \,,\\
 & \frac{\partial \,}{\partial P_\omega} \left( G_\mu S^{\mu \nu} P_\nu \right) \cong G_\mu S^{\mu \omega} \,,
\end{align}
where $F, G_\mu$ are arbitrary functions of the phase-space coordinates $x^\kappa, P_\lambda, S^{\gamma \delta}$. By choosing appropriate $F, G_\mu$, we are able to reproduce all the terms on the right hand side of \eqref{eq:hamveltd} and thus obtain the Hamiltonian
\begin{align}\label{eq:tdham}
H_\mathrm{TD} = \frac{m}{2\mathcal{M}^2}  \Bigg[ \left( g^{\mu\nu} - \frac{4 S^{\nu\gamma} R^\mu_{\; \gamma \kappa \lambda} S^{\kappa \lambda} }{4 \mathcal{M}^2 +  R_{\chi \eta \omega \xi} S^{\chi \eta} S^{\omega \xi}}  \right) P_\mu P_\nu + \mathcal{M}^2  \Bigg] \cong 0 \,,
\end{align}
where we substitute the expression \eqref{eq:mtd} for $m$. A straight-forward computation of Hamilton's equations of motion then shows that they agree with the MPD equations of motion under the TD supplementary condition \revision{given that we choose initial data such that $S^{\mu\nu}P_\nu = 0$}. 

An interesting fact discussed in Appendix \ref{app:cons} is that the Hamiltonian (for a different time parametrization, $\lambda \neq \tau$) can be obtained by applying $S^{\mu\nu}P_\nu = 0$ as a Hamiltonian constraint of the Khriplovich Hamiltonian \eqref{eq:kripham}. However, this procedure does not seem to work for any other supplementary condition. 


\subsection{The MP Hamiltonian} \label{subsec:hammp}

Similarly to the TD condition, we are now looking for a Hamiltonian that generates the MP momentum-velocity relation \eqref{eq:mppu}
\begin{align}
\frac{\partial H}{ \partial P_\mu} \cong \frac{1}{m} P^\nu\left(\delta^\mu_\nu - \frac{1}{S^2} S^{\mu\kappa}S_{\nu \kappa}\right) \,.
\end{align}
We can compose it from the single on-shell condition $P_\mu P_\nu (g^{\mu\nu} - S^{\mu\kappa}S^\nu_{\; \kappa}/S^2) = -m^2$ similarly to the previous section ($m$ is now a fixed number independent of the phase-space variables) to obtain
\begin{align}
H_\mathrm{MP} = \frac{1}{2m} \left( g^{\mu\nu} -  \frac{1}{S^2} S^{\mu\kappa}S^\nu_{\; \kappa} \right) P_\mu P_\nu\cong-\frac{m}{2} \,.
\label{eq:mpham}
\end{align}
Once again, the computation of the equations of motion shows that they are identical to the MPD equations under the MP condition. \revision{As should already be clear from the discussion in Subsection \ref{subsec:mp}, the only condition on the validity of the evolution is for the initial spin tensor to be degenerate in some time-like direction.}


\subsection{Coordinate-time parametrization} \label{subsec:tparam}

All of the above-stated Hamiltonians \eqref{eq:kripham}, \eqref{eq:tdham}, and \eqref{eq:mpham} generate motion parametrized by proper time $\tau$. It is possible to generalize them to any time parametrization $\lambda$ with $\di \lambda/ \di \tau$ an arbitrary function of any variables by exploiting the fact that the Hamiltonians have a constant value for any trajectory. We can then get the new $\lambda$-Hamiltonians as
\begin{align}
H_{\lambda} = \left( \frac{\di \lambda}{\di \tau} \right)^{-1}(H_{\tau} - H_0)\,. \label{eq:repham}
\end{align}
The constant $H_0$ is $-m/2$ for the KS and MP Hamiltonians \eqref{eq:kripham} and \eqref{eq:mpham}, and $0$ for the TD Hamiltonian \eqref{eq:tdham}. These Hamiltonians evolve the full set of variables $x^\mu,P_\nu, S^{\gamma\kappa}$.

However, it is also possible to use the component of ``non-covariant'' momentum $p_t$ from Eq.~\eqref{eq:pnonc} expressed as a function of the other variables to generate the equations of motion parametrized by coordinate time $t$. To show this in the simplest possible way, we pass to the coordinates $p_\mu, S^{AB}$ defined in equations \eqref{eq:stetr} and \eqref{eq:pnonc}. We compute
\begin{align}
\frac{\di p_i}{\di t} &= -\frac{\partial H}{\partial x^i} \left( \frac{\partial H}{\partial p_t}\right)^{-1} = - \frac{\partial(-p_t)}{\partial x^i}\Big|_{H = \mathrm{const.}} \,, \\
\frac{\di x^i}{\di t} &= \frac{\partial H}{\partial p_i} \left( \frac{\partial H}{\partial p_t}\right)^{-1} = \frac{\partial(-p_t)}{\partial p_i}\Big|_{H = \mathrm{const.}} \,, \\
\begin{split}
\frac{\di S^{AB}}{\di t} &= \{S^{AB},S^{CD}\} \frac{\partial H}{\partial S^{CD}} \left( \frac{\partial H}{\partial p_t}\right)^{-1} \\
&= \{S^{AB},S^{CD}\}\frac{\partial(-p_t)}{\partial S^{CD}}\Big|_{H = \mathrm{const.}} \,,
\end{split}
\end{align}
where we have used the implicit function theorem. In other words, for any phase-space function $F(x^i,P_i,S^{AB})$
\begin{align}
\frac{\di F}{\di t} = \{F,-p_t\Big|_{H=const.}\}\,.
\end{align}
We now list the respective Hamiltonians $H_t = -p_t|_{H=const.}$ for the KS, TD, and MP spin conditions
\begin{align}
& H_{t\rm KS} = -P_i \omega^i+ \sqrt{\alpha^2 m^2 + \gamma^{ij}P_i P_j }- \frac{1}{2}e_{\nu A;t}e^\nu_B S^{AB} \,,
\\
& \omega^i \equiv -\frac{g^{ti}}{g^{tt}}\,,\; \alpha \equiv \frac{1}{\sqrt{-g^{tt}}} \,,\;\gamma^{ij} = -\frac{g^{ij}}{g^{tt}} + \omega^i \omega^j \,,
\\
& H_{t\rm TD} = -P_i \tilde{\omega}^i+ \sqrt{\tilde{\alpha}^2\mathcal{M}^2 + \tilde{\gamma}^{ij}P_i P_j }- \frac{1}{2}e_{\nu A;t}e^\nu_B S^{AB} \,,
\\
& \tilde{g}^{\mu\nu} \equiv g^{\mu\nu} + \frac{4 S^{\gamma(\nu} R^{\mu)}_{\,\; \gamma \kappa \lambda} S^{\kappa \lambda} }{4 \mathcal{M}^2 +  R_{\chi \eta \omega \xi} S^{\chi \eta} S^{\omega \xi}} \,,
\\
& \tilde{\omega}^i \equiv -\frac{\tilde{g}^{ti}}{\tilde{g}^{tt}}\,,\; \tilde{\alpha} \equiv \frac{1}{\sqrt{-\tilde{g}^{tt}}}\,,\; \tilde{\gamma}^{ij} = -\frac{\tilde{g}^{ij}}{\tilde{g}^{tt}} + \tilde{\omega}^i \tilde{\omega}^j \,,
\\
& H_{t\rm MP} = -P_i \bar{\omega}^i+ \sqrt{\bar{\alpha}^2 m^2 + \bar{\gamma}^{ij}P_i P_j }- \frac{1}{2}e_{\nu A;t}e^\nu_B S^{AB} \,,
\\
& \bar{g}^{\mu \nu} \equiv g^{\mu\nu} -  \frac{1}{S^2} S^{\mu\kappa}S^\nu_{\; \kappa} \,,
\\
& \bar{\omega}^i \equiv -\frac{\bar{g}^{ti}}{\bar{g}^{tt}}\,,\; \bar{\alpha} \equiv \frac{1}{\sqrt{-\bar{g}^{tt}}}\,,\; \bar{\gamma}^{ij} = -\frac{\bar{g}^{ij}}{\bar{g}^{tt}} + \bar{\omega}^i \bar{\omega}^j \,,
\end{align}
where we have chosen roots of $p_t$ corresponding to particles traveling forward in time. 

Now we have to choose a set of phase-space coordinates to be evolved by the Hamiltonians above. First possible set of variables, to be evolved with the use of the spatial part of the Poisson brackets \eqref{eq:poiss}, is $x^i,P_j,S^{\mu\kappa}$, where we have to use $S^{AB} = S^{\mu\nu}e_\mu^A e_\nu^B$.\footnote{Note that in stationary space-times and for time-independent tetrads $e_{\kappa A;t}e^\kappa_B e_\mu^A e_\nu^B = \Gamma_{[\mu\nu] t}$, and in this case the Hamiltonians in the $x^i,P_j,S^{\mu\kappa}$ basis will be tetrad-independent.} A second option is to transform all variables into $x^i, p_j, S^{AB}$ with the use of the orthonormal tetrad basis $e_A^\mu$, as given in equations \eqref{eq:partcan}. The resulting Hamiltonians are to be evolved with the use of the spatial part of the brackets \eqref{eq:partcan}.

\revision{The Hamiltonians above should be compared to the coordinate-time Hamiltonian for the NW condition given by \citet{barausse2009}. Our expressions differ in the fact that they use an $\mathrm{SO}(3,1)$ rather than an $\mathrm{SO}(3)$ commutation structure for the spin, and the equations are under different supplementary conditions. In return, our Hamiltonians reproduce the pole-dipole MPD equations \eqref{eq:mpd} exactly, with no truncations in powers of spin (other than those intrinsic to the pole-dipole approximation itself).}


\section{Canonical coordinates and numerical integration} \label{sec:canon}

In this Section, we elaborate on the structure of the phase space.
Being a geometric space (symplectic manifold), constraints can be viewed as defining a
surface/submanifold. \revision{One can (but does not have to) project} the Poisson bracket into the constraint
surface, which leads to the so-called Dirac bracket \cite{dirac1966,hanson1976}. Furthermore,
canonical coordinates can be adopted (at least locally), which we construct explicitly.
This is crucial for the symplectic integration studied in the next Section.


\subsection{Importance of canonical coordinates and Dirac brackets}\label{subsec:whycan}

Let us assume that we have a set of constraints $\Phi^a=0$ and the constraint algebra $C^{ab} = \{\Phi^a,\Phi^b\}$ with $C^{ab}$ a non-degenerate matrix with an inverse $C_{ab}^{-1}$. Then it is possible to define a new constrained Poisson bracket \citep{dirac1966,hanson1976}
\begin{align}
\{A,B\}' = \{A,B\} - \{A,\Phi^a\}C^{-1}_{ab}\{\Phi^b,B\} \,.
\end{align}
The bracket $\{,\}'$ is often called the Dirac or Dirac-Poisson bracket. If we have a Hamiltonian that fulfills $\{\Phi^a,H\} \cong 0$, then the equations of motion generated by $\{,\}'$ and $H$ are the same as with $\{,\}$ and $H$. The bracket-constraining procedure was originally devised for the purposes of canonical quantization. Nonetheless, it is also useful for classical Hamiltonian dynamics. 

When we want to study a classical Hamiltonian system at high accuracy over a large number of periods (such as would be the case of EMRIs), it is highly advantageous to use symplectic integration \citep[see e.g.][]{hairer2006}. Most symplectic integrators require that the equations are formulated in terms of pairs of canonical coordinates, i.e. a colection of phase-space coordinates $\chi^i,\pi_i$, with $i$ some labelling index, such that $\{\chi^i,\pi_j\}=\delta^i_j$ (however, there do exist symplectic integrators for special classes of systems that require no such coordinates \citep{mclachlan2014,mclachlan2017}). 

The usefulness of the constrained bracket $\{,\}'$ in this context can be twofold. First, it may be easier to find canonical coordinates for $\{,\}'$ rather than $\{,\}$. Second, the constraints $\Phi^a =0$ are only integrals of motion with respect to the dynamical system evolved by the unconstrained bracket $\{,\}$, and they cannot be forced to be zero during integration, otherwise the advantageous properties of the symplectic algorithm are broken. On the other hand, in the case of the bracket $\{,\}'$, the constraints $\Phi^a$ commute with any phase-space variable. In return, they are effectively promoted to a ``phase-space identity'' and can be used to reduce the number of variables in a numerical integrator symplectic with respect to $\{,\}'$.  

For example, \citet{barausse2009} applied the NW supplementary condition as a constraint to the bracket \eqref{eq:partcan} (along with brackets and constraints for auxilliary variables) to obtain, at least at linear order in spin, a simplified bracket for the reduced number of variables $p_\mu,x^\nu, S^{IJ}$ (see Subsection \ref{subsec:cpnw}). This system was then easy to cover by approximate canonical coordinates and thus to study by symplectic integration \citep{lukes2014}.

As for the possibility to reduce the variables in the case of other supplementary conditions, the TD condition $S^{\mu\nu}P_\nu =0$ applied as a constraint leads to a very complicated Dirac bracket that mixes the spin and momentum degrees of freedom. As a result, it is very difficult to find the canonical coordinate basis for the TD-constrained bracket. 

On the other hand, as discussed in Subsections \ref{subsec:ks} and \ref{subsec:mp}, the KS and MP condition in fact do not allow to reduce the number of evolved variables to the same extent as the TD and NW/CP conditions. A closer inspection shows that the KS and MP conditions cannot even be formulated as a constraint on the phase space $p_{\mu},x^\nu,S^{AB}$, and the Poisson bracket will thus always be \eqref{eq:partcan}. Hence, for the purposes of the TD, KS, and MP conditions we have decided to find the canonical coordinates covering the full phase-space $p_{\mu},x^\nu,S^{AB}$ for the unconstrained bracket \eqref{eq:partcan}.

\subsection{Canonical coordinates on $S^{AB}$}
The $p_\mu, x^\nu$ sector of the phase-space coordinates is already canonical, so we are looking for canonical coordinates covering the spin tensor $S^{AB}$. To find the canonical coordinates, we mimic the procedure of \citet{tessmer2013} by expressing $S^{AB}$ as a simple constant tensor $S^{\hat{A}\hat{B}}$ in some ``body-fixed frame'' plus a Lorentz transformation $\Lambda^A_{\;\; \hat{A}}$ into the ``background frame'' $e^A_\mu$. The parameters of the transformation, when chosen appropriately, then turn out to be canonically conjugate pairs of coordinates.  

The details of the procedure are given in Appendix \ref{app:canon}, we only summarize here the resulting coordinates
\begin{subequations}
\label{eq:spcanon}
\begin{align}
&A = S^{12} - \sqrt{(S^{12})^2 + (S^{23})^2 + (S^{31})^2} \,,\\
&B = \sqrt{(S^{12})^2 + (S^{23})^2 + (S^{31})^2} - S\,,\\
&\phi = \arctan\left( \frac{S^{31}}{S^{23}} \right) \,,\\
&\psi = \arctan\left( \frac{S^{31}}{S^{23}} \right) -\kappa \arccos\left( S^{03}\sqrt{\mathcal{C}} \right) \,,\\
&\kappa = \mathrm{sign}(S^{02} S^{23} - S^{01} S^{31}) \,,\\
&\mathcal C = \frac{(S^{12})^2 + (S^{23})^2 + (S^{31})^2}{\left[(S^{31})^2 + (S^{23})^2\right] \left[(S^{01})^2 + (S^{02})^2 + (S^{03})^2\right]} \,.
\end{align}
\end{subequations}
Even though the construction in Appendix \ref{app:canon} provides the path to the derivation of these coordinates, one may simply verify their Poisson brackets by direct computation. The brackets then are $\{\phi,A\} = \{\psi,B\} = 1$ and $0$ otherwise. 

The backwards transformations from the canonical coordinates to the spin tensor read 
\begin{subequations}
\label{eq:spparam}
\begin{align}
&S^{01} = \mathcal{D} \left[ A \cos(2 \phi - \psi) + (A + 2 B + 2S) \cos \psi \right] \,,\\
&S^{02} = \mathcal{D} \left[ A \sin(2 \phi - \psi) + (A + 2 B + 2S) \sin \psi \right] \,,\\
&S^{03} = - 2\mathcal{D}\mathcal{E}  \cos(\phi -\psi) \,,\\
&S^{12} = A+B+S\,,\\
&S^{23} = \mathcal{E} \cos\phi\,,\\
&S^{31} = \mathcal{E} \sin\phi\,,\\
& \mathcal{D} =  -\frac{\sqrt{B(B + 2S)}}{2(B+S)} \,, \\
& \mathcal{E} =  \sqrt{-A(A+2B + 2S)} \,. 
\end{align}
\end{subequations}
The coordinates cover the space of general antisymmetric tensors with a degenerate time-like direction and a closer consideration reveals a number of similarities with hyperspherical coordinates in $\mathbb{R}^4$. 

The coordinates have singularities at $B=0$ and $A=0,-2(B+S)$ which have the character similar to those of the singularities at $r=0$ and $\cos(\vartheta) =1,-1$ in spherical coordinates in $\mathbb{R}^3$. \revision{Another formal singularity appears in the definition of $\psi$ at $S^{02}S^{23} - S^{01} S^{31} = 0$. This occurs only due to the necessity of switching between the branches $\kappa = \pm 1$ of $\arccos()$ and its formal nature can be seen, for example, from the smoothness of the inverse parametrization \eqref{eq:spparam} around the respective points.}

As a result, the physical coordinate ranges then are $B\in(0,  \infty)$ and $A\in(-2(B+S),0)$. The coordinates $\phi,\psi$ are simple angular coordinates similar to the azimuthal angle $\varphi$ in spherical coordinates in $\mathbb{R}^3$, and they both run in the $[0,2\uppi)$ interval. Some more details about the coordinate singularities are given in Appendix \ref{app:canon}.

One last remark is that the coordinates $\phi,\psi$ are dimensionless and have finite limits as $S \to 0$, whereas $A,B$ have the dimension of the spin and should generally go to zero when $S \to 0$. However, if we keep $a\equiv A/S,b\equiv B/S$ finite, then the evolution of the coordinates $a,b,\phi,\psi$ can be used to track the evolution of a ``test spin'', i.e. an intrinsic spin of the particle that is transported along the trajectory while not exerting any back-reaction on the orbit itself.

There is a special case when $A,B$ can remain finite while $S \to 0$, and that corresponds to the body-fixed frame being infinitely boosted with respect to the background frame and the vanishing direction of the spin tensor becoming light-like. This particular limit may be useful for the description of massless particles with spin but we consider it to be physically meaningless for the current context of massive bodies.


\section{Special planar motion} \label{sec:plan}


We now want to study a simple restricted problem that would allow us to demonstrate the practical properties of the canonical coordinates. We do so by considering a motion in the equatorial plane of the Schwarzschild space-time under the KS condition. Then we require that both the four-velocity and the spin tensor are initially vanishing in the $\vartheta$ direction, $S^{\mu \vartheta} = \dot{\vartheta} = 0$. We then easily compute that
\begin{align}
&\frac{\di^2 \vartheta}{\di \tau^2} = 0 \,,\\
&\frac{\di S^{\mu \vartheta}}{\di \tau} = 0 \,.
\end{align}
In other words, the conditions $S^{\mu \vartheta} = 0,\,\dot{\vartheta} = 0$ will be satisfied throughout the motion. 

A similar system restricted to the equatorial plane can be formulated by requiring $P_\vartheta = S^{\mu \vartheta}=0$ also for the MP and TD conditions, and, furthermore, the background could be generalized to the Kerr space-time. However, we choose here to study the special planar problem only in the KS incarnation and in the Schwarzschild space-time because of its simplicity. 

\revision{As can be seen from the discussion in Subsection \ref{subsec:ks}, the freedom in the initial conditions we have in the spin sector (once the orbital initial conditions have been chosen) in the case of the KS condition is of a gauge nature; it corresponds to the freedom in choosing an initial moment-defining frame $w^\mu$. In the case of a strictly observationally driven study, we would, quite naturally, restrict such choices of $w^\mu$. Nevertheless, since we are mainly concerned here with demonstrating the power of the canonical coordinates in a numerical integration, we treat this additional freedom simply as another phase-space variable.}


\subsection{The Hamiltonian}

For our computations, we choose the coordinate-aligned tetrad in the usual Schwarzschild coordinates $t,\varphi,r,\vartheta$: $e^0_\mu = \sqrt{-g_{tt}} \delta^t_\mu, \, e^1_\mu = \sqrt{g_{\varphi\varphi}} \delta^\varphi_\mu, \, e^2_\mu = \sqrt{g_{rr}} \delta^r_\mu, \, e^3_\mu = \sqrt{g_{\vartheta \vartheta}} \delta^\vartheta_\mu$. The choice of the tetrad and even the order of the legs are important for the final form of the Hamiltonian and the physical interpretation of the quantities appearing in it. However, the choice of the tetrad never matters for the real physical evolution of the KS, TD, or MP conditions (unlike in the case of the NW/CP condition where the choice $\sim \xi^\mu \sim e^\mu_0$ is crucial \citep{kunst2016}).

The condition $S^{\vartheta \mu}=0$ then translates into either $A = 0$ or $A = -2(B+S)$ for $S^{12} >0$ and $S^{12} <0$ respectively. Here we choose $S^{12}>0$, and $\phi$ thus becomes a redundant coordinate (see more details in Appendix \ref{app:canon}). In a typical right-hand-oriented interpretation and for an orbit with positive $\dot{\varphi}$, this corresponds to a spin vector counter-aligned to the orbital angular-momentum vector.

When the dust settles, the Hamiltonian \eqref{eq:kripham} expressed in canonical coordinates in the case of the special planar motion reads
\begin{align}
H_\mathrm{SP} &= \frac{1}{2m}\Bigg[ \frac{-1}{1-2M/r} \left( p_t - \frac{M \sqrt{B(B+2S)}\sin \psi}{r^2}\right)^2 \label{eq:hsp}\\ \nonumber &+ \left(1 - \frac{2M}{r}\right) p_r^2 + \frac{1}{r^2}\left(p_\varphi - \sqrt{1 - \frac{2M}{r}} (B+S)\right)^2 \Bigg] \,.
\end{align}
The system has two obvious integrals of motion $p_\varphi,p_t$, since the coordinates $t,\varphi$ are cyclic. However, it should be noted that the orbital angular momentum and energy will generally vary during the evolution since they relate to the phase-space coordinates as
 \begin{align}
 & u_t = \frac{1}{m} \left(p_t - \frac{M \sqrt{B(B+2S)}\sin \psi}{r^2} \right) \,, \label{eq:utsp}\\
 & u_\varphi = \frac{1}{m} \left(p_\varphi - \frac{r^{5/2}(B+S)}{\sqrt{r-2M}} \right)\, \label{eq:uphsp}.
\end{align}

\subsection{Poincar{\'e} surfaces of section}

\begin{figure*}
\centering
\begin{minipage}{1\textwidth}
\includegraphics[width=0.49\linewidth]{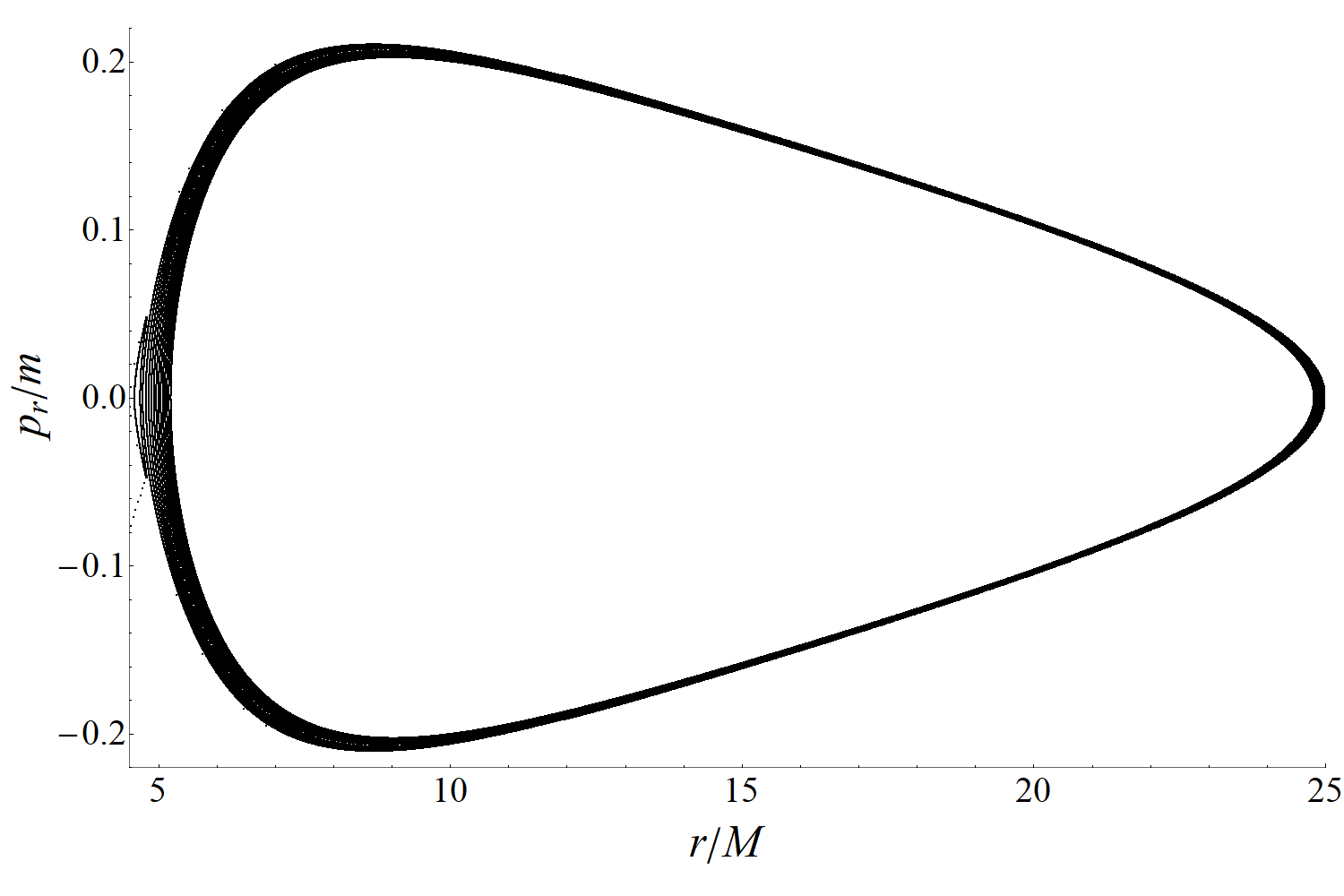}
\includegraphics[width=0.49\linewidth]{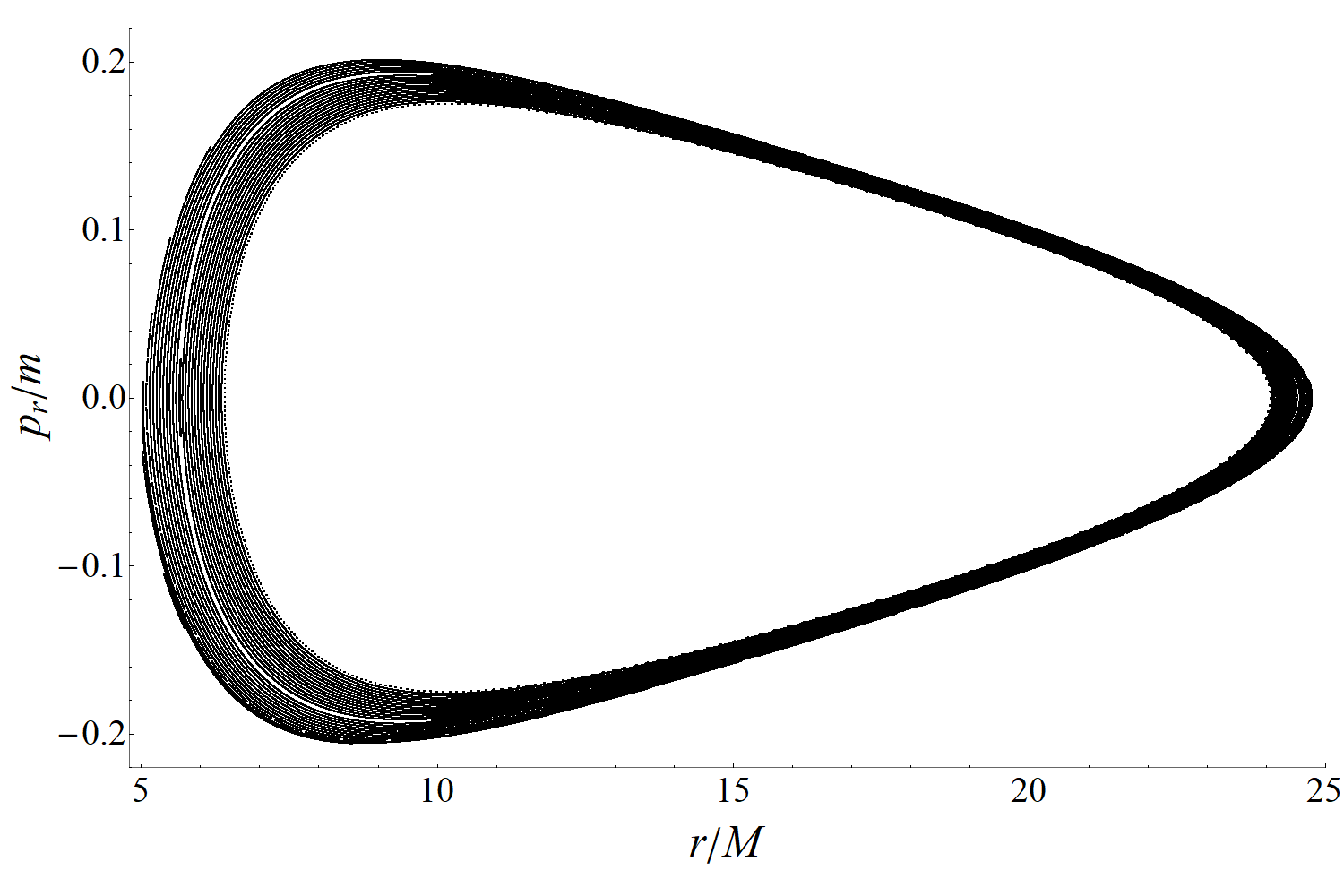} \\
\includegraphics[width=0.49\linewidth]{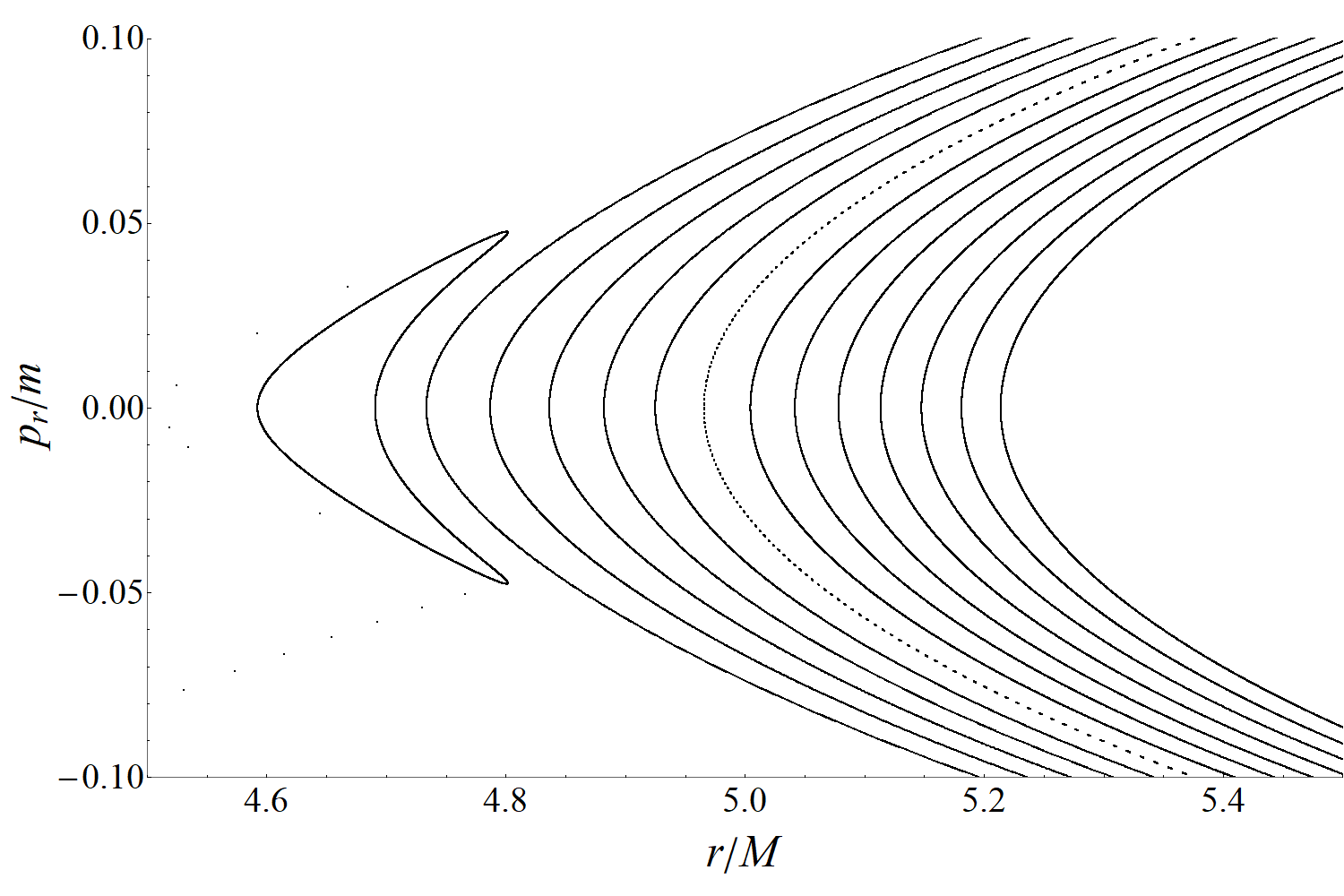}
\includegraphics[width=0.49\linewidth]{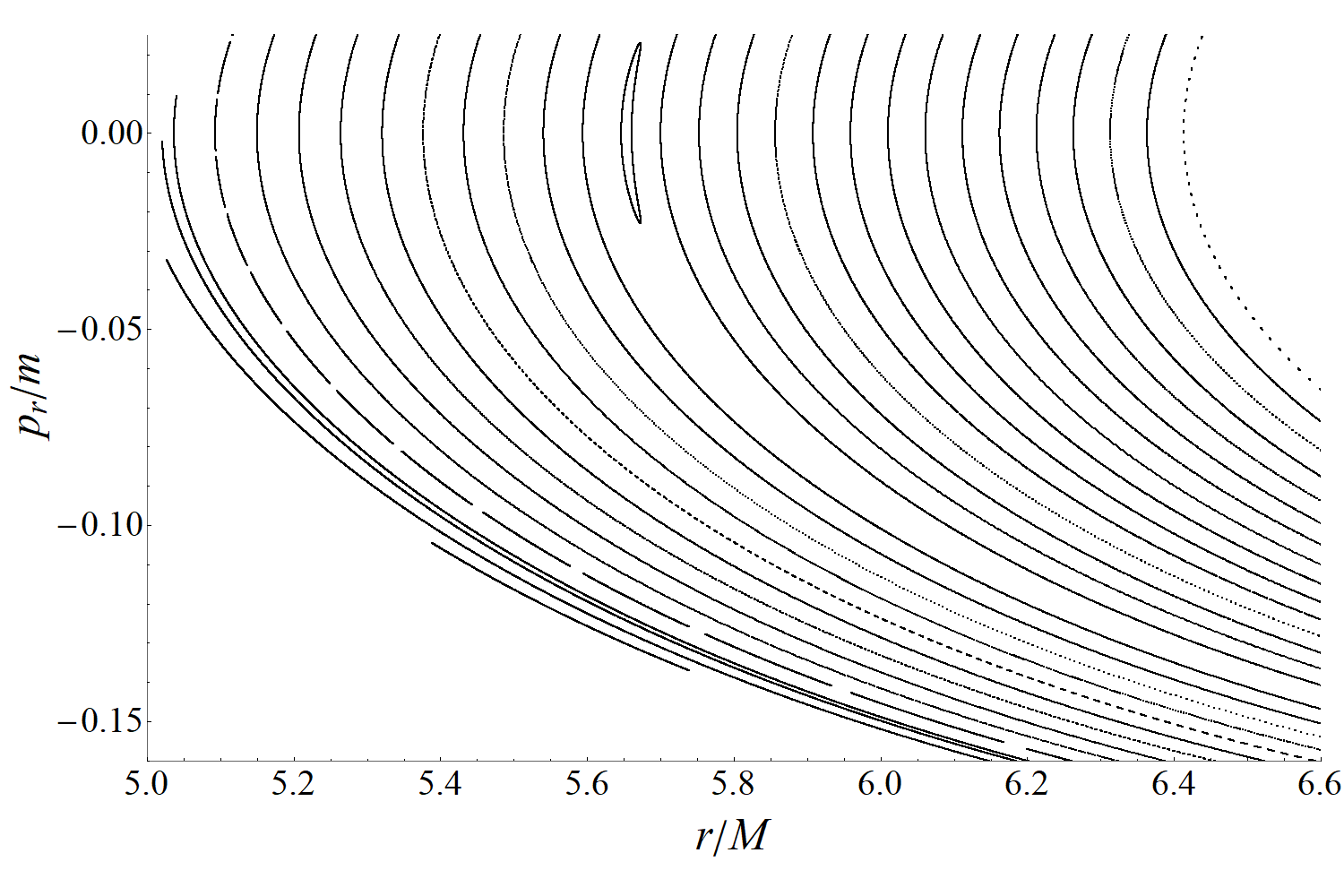} \\
\includegraphics[width=0.49\linewidth]{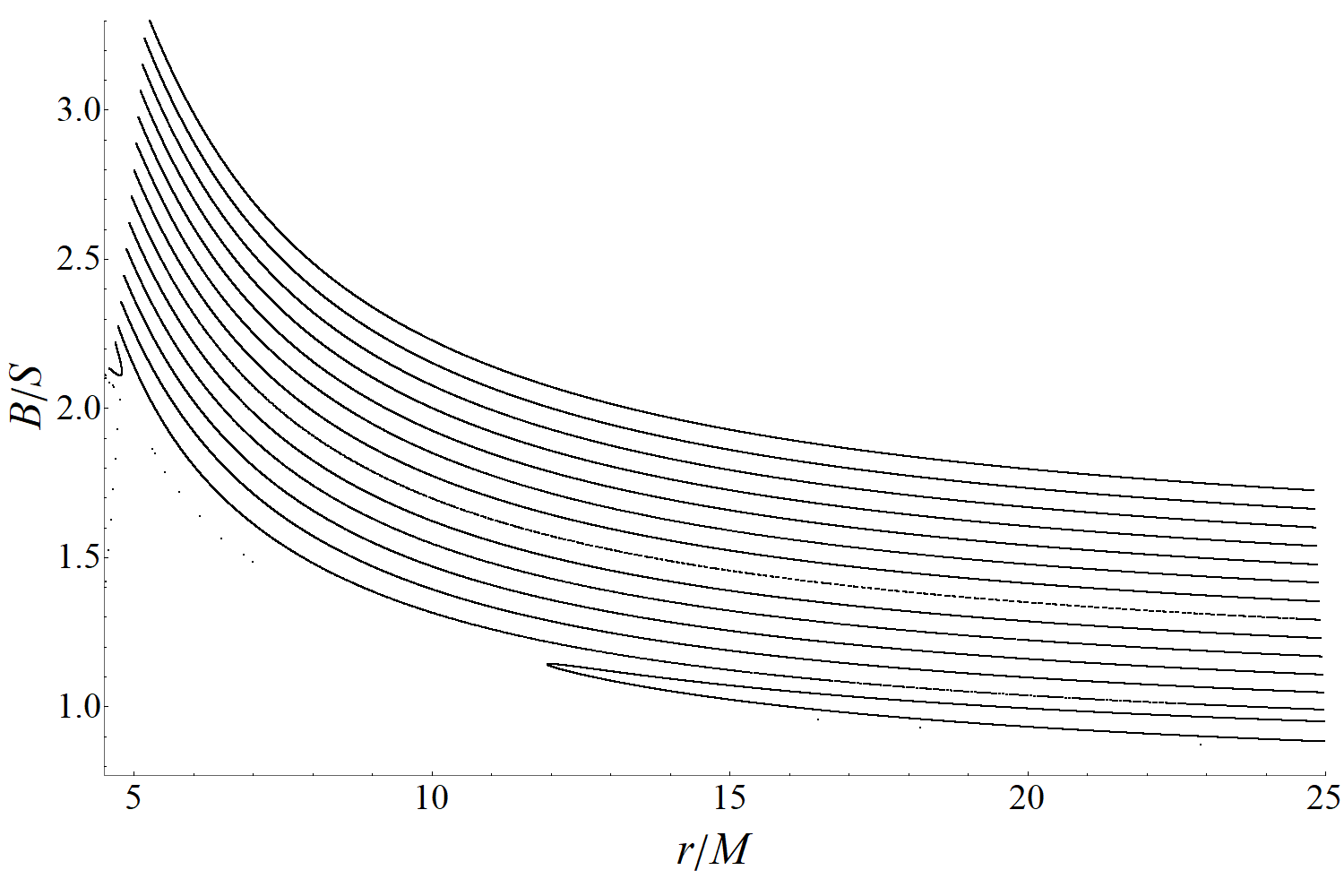}
\includegraphics[width=0.49\linewidth]{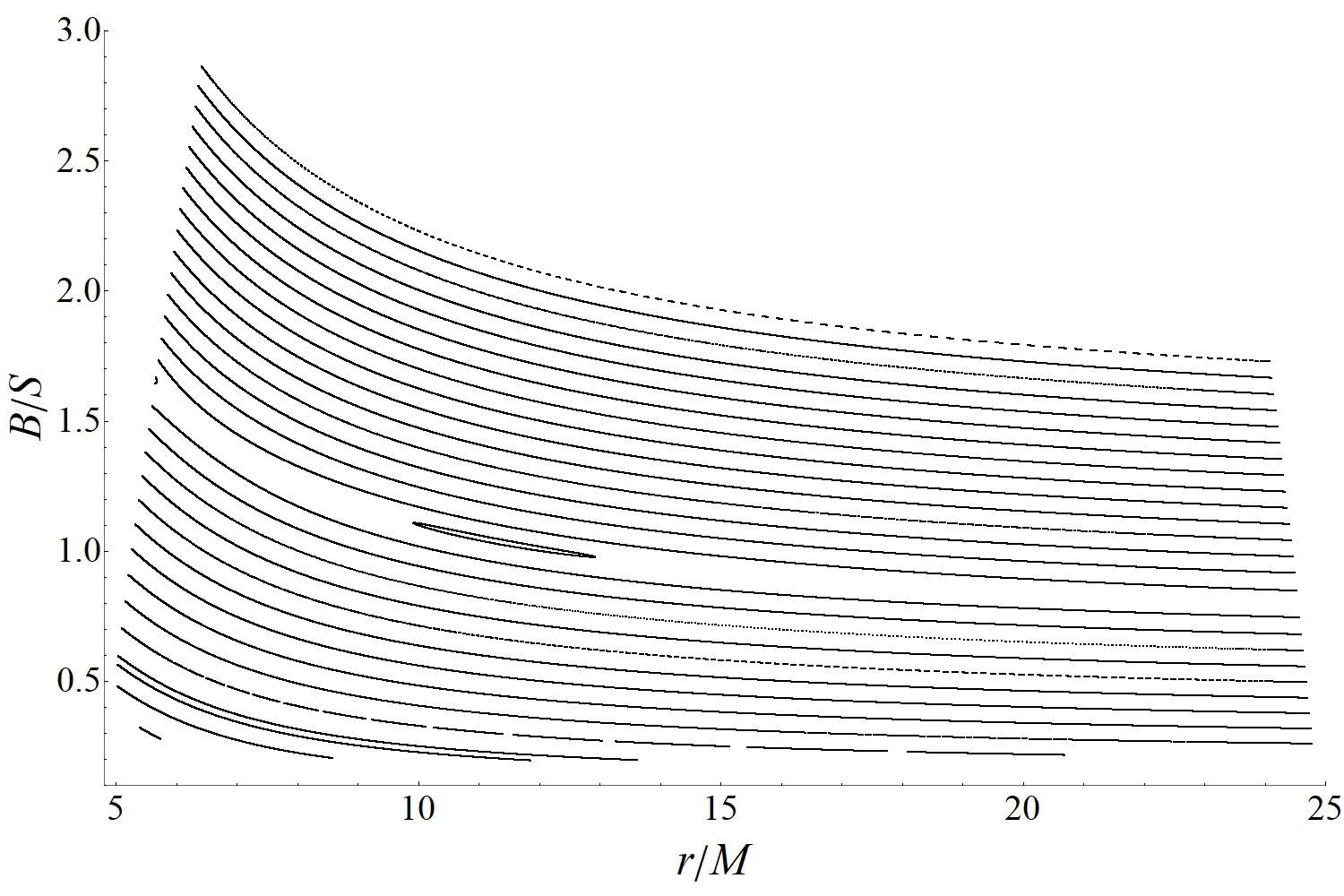} \\
\end{minipage}
\caption{Poincar{\'e} surfaces of section for the special planar problem at $p_t/m = -0.97, p_\varphi/m = 3.7 M$ created by snapshots after every cycle in the spin-angle $\psi$. The left column corresponds to $S/m = 0.05M$ and the right column to $S/m = 0.1M$. The outer parts of the nested sections correspond to small $B/S$ whereas the inner parts to growing $B/S$.  The left column features a smaller number of orbits because the ``outer'' orbits are plunging into the black hole.} \label{fig:sections}
\end{figure*}

\begin{figure*}
\centering
\begin{minipage}{1\textwidth}
\includegraphics[width=0.49\linewidth]{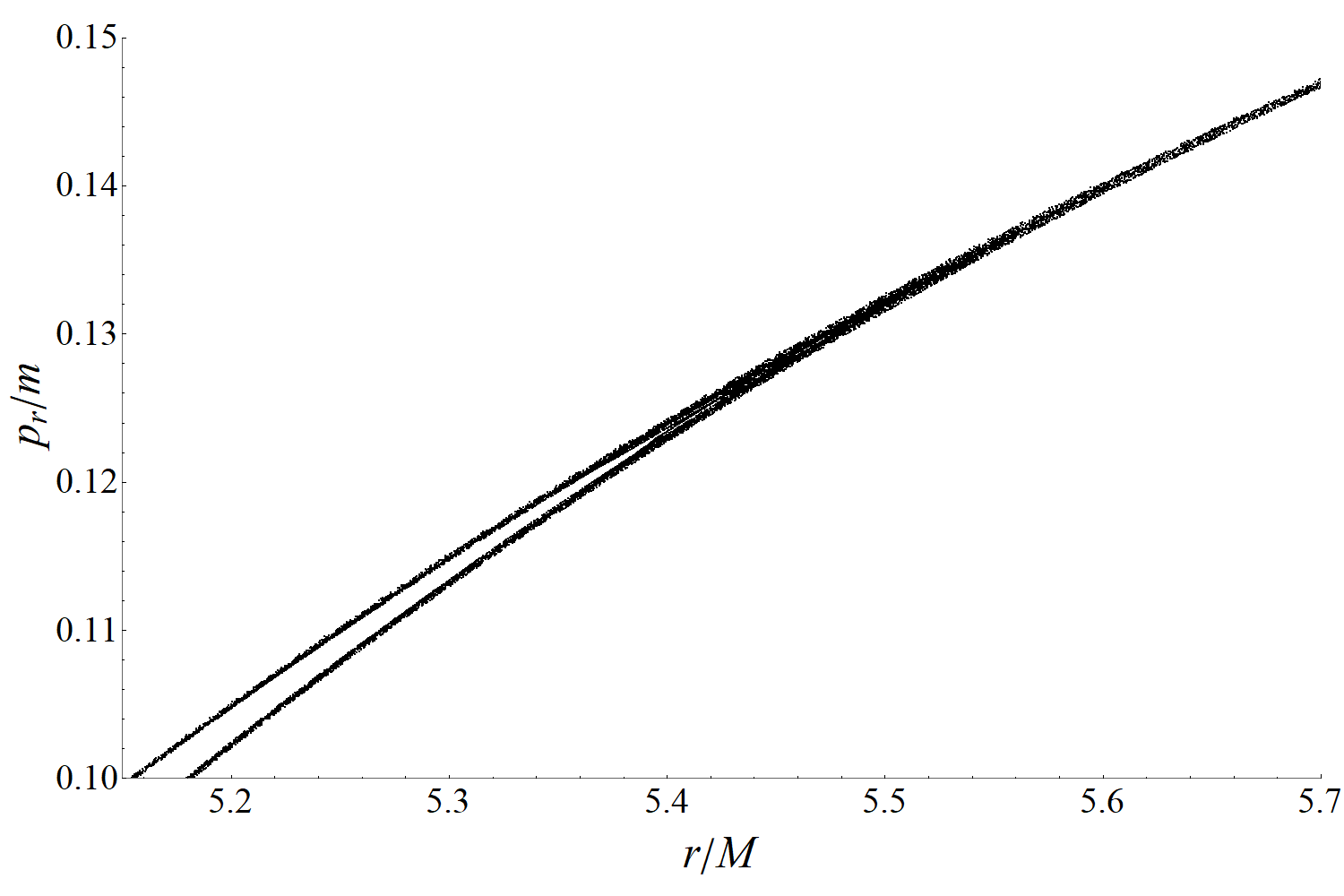}
\includegraphics[width=0.49\linewidth]{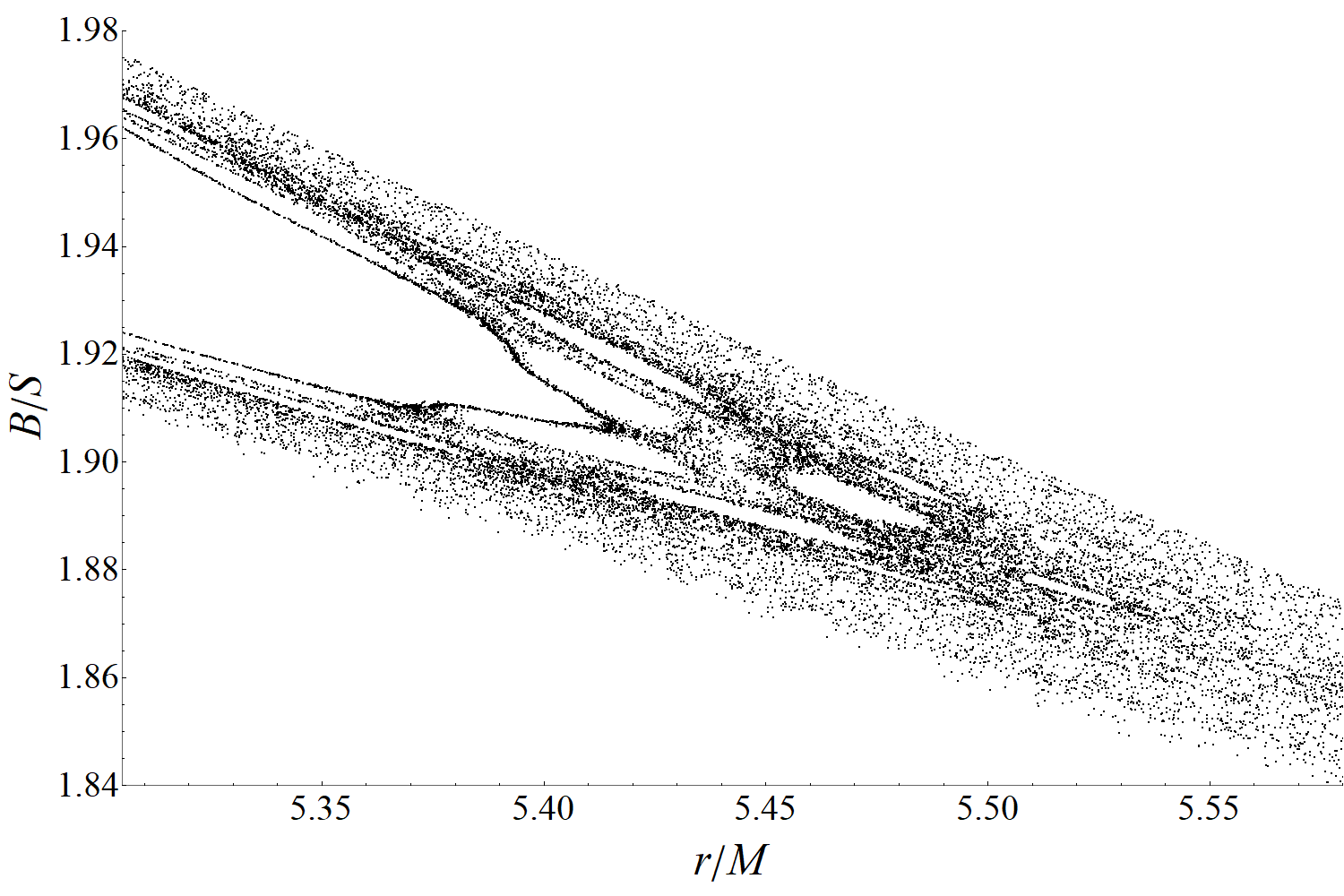} \\
\end{minipage}
\caption{A detailed Poincar{\'e} surface of section of a single chaotic trajectory at $p_t/m = -0.97, p_\varphi/m = 3.7 M, S/m = 0.05M$ (compare with left column of Fig. \ref{fig:sections}). The trajectory was integrated over $\sim 10^5$ spin cycles ($\sim 10^4$ orbital cycles) so that we would observe its release from the sticky fractal layer around the resonance into the general separatrix chaos.} \label{fig:chaos}
\end{figure*}

We have constructed the problem so that only two degrees of freedom become dynamically important, $r, p_r$, and $\psi,B$. Since the trajectory is also constrained by four-velocity normalization, all the phase-space trajectories of a given $p_\varphi, p_t$ are then confined to a 3-dimensional hypersurface. We make a natural Poincar{\'e} surface of section through this hypersurface by sampling this set of trajectories and recording the phase-space variables every time $\psi$ finishes a $2 \uppi$ cycle. Thanks to this construction, we obtain well-defined 2D Poincar{\'e} surface of section, whereas in the general case the surface of section becomes higher-dimensional and new methods need to be employed for visualization \citep[see][]{lukes2016}.

We have integrated the trajectories using the 6-th order Gauss collocation scheme with a fixed-point iteration of the collocation points \citep[see, e.g.,][]{hairer2006}. Additionally, we exploited the parametrization invariance of the trajectory by using a time parameter $\lambda$ such that
\begin{equation}
\frac{\di \lambda}{\di \tau} = \frac{r_0^2}{r(r-2M)} \frac{ S}{\sqrt{B(B+2S)} + \epsilon}\,,
\end{equation}
where $\epsilon, r_0$ are constants we set to $10^{-4},10M$ respectively. This effective time-stepping does not spoil the symplecticity of the integrator because the respective equations of motion can be generated by a Hamiltonian of the form \eqref{eq:repham}.

With these measures in place and by using a standard single-thread computation in C++, we were able to integrate through $10^4$ spin cycles within minutes at a relative error less than $10^{-12}$ in the four-velocity normalization. Of course, such efficiency and long-term accuracy would hardly be possible without the canonical coordinates and geometrical integration. This is an important point of the present section.

In general, the dimensionless parameter $S/(Mm)$ can be understood as a perturbation strength non-linearly coupling two exactly integrable systems, the geodesic motion in Schwarzschild space-time, and the parallel transport of the ``test spin'' on top of that geodesic. As such, the dependence of the phase portrait of the special planar problem on the particle spin should have the same characteristics as any weakly non-integrable system \citep[e.g.][]{arnold2007}: The originally smooth phase-space foliation by regular oscillations of the trajectories should now feature occasional ``breaks'' in the form of resonances and thin chaotic layers.

In order to demonstrate the presence of such structures, we probed values of $p_t,p_\varphi$ so that phase-space trajectories in the studied congruence are near unstable circular orbits in the original geodesic flow. This is because the neighboring phase space also contains the ``homoclinic'' infinite whirl-zoom orbits that are well known to act as ``seeds of chaos'' in perturbed black hole space-times \citep[see, e.g.,][]{chaospapI,chaospapII,chaospapIII,chaospapIV}. However, as can be seen from equations \eqref{eq:utsp} and \eqref{eq:uphsp}, the variations of $S$ also have the unfortunate effect of shifting the meaning of $p_t,p_\varphi$, and this easily pushes us into the phase-space regions of orbits plunging into the black hole.

As already discussed in the Introduction, we should be imposing self-force effects along with the spin-force even for the smallest values of $S/(mM)$ in a self-consistent physical model. Thus, it does not make sense to study the influence of the spin beyond perturbation-like values and we choose to study only $S/(mM) \leq 0.1$.  

Furthermore, the ratio $B/S$ is equal to $\gamma -1$, where $\gamma$ is the usual gamma factor of the Lorentz boost from the body-fixed frame to the background frame (see Appendix \ref{app:canon}) \revision{and we will have $(S^{\mu\nu}\dot{x}_\nu)/S \propto B/S$. If we compare this with the discussion in Subsection 2.1, we see that if $B/S$ becomes too large, the world-line becomes shifted outside of the interior of the real physical body, and the system of equations instead obtains the character of some sort of perturbed geodesic-deviation equation.} Hence, we only allow $B/S \lesssim 5$ in our initial conditions.

We show two Poincar{\'e} surfaces of section in the relevant ranges in Fig. \ref{fig:sections}. In these sections, we are able to find resonances corresponding to ratios as low as $1:1$ or $1:2$ in the spin-orbital frequencies. Additionally, small chaotic layers can be found near the saddle points of the resonant chains (see Fig. \ref{fig:chaos}). 

As we go to smaller values of $S/(mM)\lesssim 0.01$, the resonances become extremely thin, and most of the chaotic structure is disqualified based on the criterion $B/S \lesssim 5$. If we ignore the $B/S$ criterion and go to $B/S \gtrsim 10$, chaos can be found up to $S/(mM) \sim 10^{-3}$.

\subsection{Comparison with previous results}

Let us briefly compare these results with the study of \citet{lukes2016}, who studied the chaotization of general orbits of spinning particles in Kerr space-time while using the NW-condition Hamiltonian of \citet{barausse2009}. The motion of non-planar orbits with general spin orientations leads to richer dynamics, as an additional \revision{physical} degree of freedom enters the interactions. Consequently, Lukes-Gerakopoulos et al. found chaotic motion in the phase space until $S/(mM) = 10^{-3}$. 

As discussed in Subsection \ref{subsec:cpnw}, the NW condition constrains one more degree of freedom than the KS condition. Consequently, an analogous special planar motion $P^\vartheta = \dot{x}^\vartheta = S^{\mu \vartheta} = 0$ would in fact have only a single active degree of freedom under the NW condition and would thus be integrable at any value of spin. Additionally, even the non-planar motion under the NW Hamiltonian of \citet{barausse2009} is integrable to linear order in $S$ in Schwarzschild space-time, at least under the right choice of $\xi^\mu$ \citep{kunst2016}. \revisiontwo{Hence, the chaos found in our study using the KS condition is a qualitatively different feature as compared to a treatment using the NW condition.}

\revision{The pole-dipole MPD equations under different supplementary conditions are equivalent to each other only to dipole order and the transformation to another supplementary condition brings about shifts in the quadrupole and higher-order momenta. In this sense, it would seem that the chaos appearing in our study is a consequence of higher-order effects and/or of stretching of the physicality of the KS condition in the sense as discussed in Subsection \ref{subsec:ks}. On the other hand, the semi-linear form of the KS+MPD system of equations \eqref{eq:ks} would invite speculations that they are most likely to be exactly integrable in simple space-times. Our work provides a clear counter-example to such mathematical conjectures, the system \eqref{eq:ks} is not integrable in Schwarzschild space-time even in the planar case.}  Hence, one should be cautious in issuing general statements about the (non)-integrability and chaos in MPD equations near black holes, because such statements seem to be dependent on the context and approximations made.

\section{Conclusions and outlooks}

In this paper we have built the Hamiltonian formalism for spinning particles under all commonly used ``comoving'' supplementary conditions, that is, conditions that utilize only the local dynamics of the body and no background vector field. The full set of canonical coordinates that we provide is the minimal set of variables needed to evolve the Mathisson-Pirani and Kyrian-Semer{\'a}k conditions. Hence, our formalism allows to integrate the respective equations at peak efficiency. However, the canonical coordinates contain a redundant degree of freedom for the case of the Tulczyjew-Dixon condition. Nevertheless, a pair of extra variables to be evolved in a numerical routine is a small price to pay for the long-term quality of the evolution such as the one seen in Section \ref{sec:plan}.

Of course, it would be interesting to see whether a minimal set of canonical coordinates can be found for the Tulczyjew-Dixon condition. In principle, this can be achieved by constraining the Poisson bracket similarly to \citet{barausse2009}, and by finding the canonical basis thereof. However, the constraint procedure introduces non-zero commutation relations between momenta, space-time coordinates, and spin degrees of freedom. Consequently, the canonical coordinates would in fact be an intricate transformation of all $p_\mu,x^\mu, S^{AB}$. We plan to investigate this possibility in future work. 

Another less obvious application of the canonical coordinates is the fact that now we are able to formulate a Hamilton-Jacobi equation by making the action $\mathcal{S}$ also a function of the spin angles $\psi,\phi$ with the gradients defining the conjugate momenta $\mathcal{S}_{,\phi} = A,\, \mathcal{S}_{, \psi} = B$. We are currently preparing a manuscript presenting solutions to the Hamilton-Jacobi equation in black hole space-times.

Similarly, we believe that the herein presented formalism can be very useful to the various averaging and two-timescale approaches to EMRIs \citep{hinderer2008,pound2008,mino2008,van2018} since we can easily construct action-angle coordinates in the spin sector and thus provide an elegant treatment of the non-dissipative (``fast'') part of the dynamics of the binary. Furthermore, the simplicity of the Hamiltonian under the Kyrian-Semer{\'a}k condition makes it an attractive alternative to the Hamiltonian of \citet{barausse2009} in EOB models \cite{barausse2010}.

\section*{Acknowledgments}
V.W. is grateful for Ph.D. supervision by Claus L{\"a}mmerzahl, and for the support from a Ph.D. grant of the German Research Foundation (DFG) within Research Training Group 1620 ``Models of Gravity''.
G.L-G and V.W. are supported by Grant No. GACR-17-06962Y of the Czech Science Foundation,
which is acknowledged gratefully. We would like to thank Dirk Puetzfeld, Justin Vines, Old{\v r}ich Semer{\'a}k, Domenico Giullini, and Robert McLachlan for valuable discussions on these topics. V.W. would also like to express gratitude for the hospitability of the Astrophysical and Cosmological Relativity group at the Albert Einstein institute in Potsdam-Golm.

\newcommand{\newblock}{}
\bibliographystyle{apalike}
\bibliography{literatura}

\clearpage
\appendix

\section{The generalized KS conditions} \label{app:ks}
The only condition that we need to be fulfilled for $\dot{S}^{\kappa \lambda} =0$ to hold is that $\dot{S}^{\mu \nu} w_\nu = 0$ for some time-like $w^\nu$. From equation \eqref{eq:scykl} projected into $w^\nu$ we then get
\begin{align}
w^\nu \dot{x}_\nu \dot{S}^{\kappa \lambda} = 0 \,.
\end{align}
Because the product of any two time-like vectors is non-zero, we then get simply $\dot{S}^{\kappa \lambda} = 0$. The supplementary condition can thus be of the form $S^{\mu\nu} w_\nu = m^\mu$ with
\begin{align}
S^{\mu\nu} \dot{w}_\nu = \dot{m}^\mu \,, \label{eq:mw}
\end{align}
because then we will have $\dot{S}^{\mu \nu} w_\nu = 0$. In the case when $m^\mu = 0$, we get that $\dot{w}_\nu$ must lay in the degenerate directions of the spin tensor, $\dot{w}_\nu = \alpha w_\nu + \beta s_\nu$ with $\alpha,\beta$ arbitrary functions of any variables \citep{semerak2015}.
However, we may generally set $m^\mu \neq 0$ and then the only condition on the evolution is Eq.~\eqref{eq:mw}. One particular option is $\dot{w}^\mu = \dot{m}^\mu = 0$. 

Nevertheless, it should be noted that only the initial choices of $w^\mu, m^\nu$ matter. This can be seen from the fact that if the equations of motion are expressed in terms of $S^{\mu\nu}$, we need no reference to $\dot{w}^\mu, \dot{m}^\nu$ as long as equation \eqref{eq:mw} is satisfied.

In summary, once we allow for $m^\mu\neq 0$, the initial conditions for $S^{\mu\nu}$ are completely unconstrained. The study of \citet{ambrosi2016} can be understood as conducted exactly in the $m^\mu\neq 0$ generalized KS condition.

One last note is that the vector $m^\mu$ represents a mass dipole in the frame $ w^\mu$, and by setting its dynamics to fulfill different evolution equations than in Eq.~\eqref{eq:mw}, we can in fact obtain other supplementary conditions \citep{ohashi2003}.


\section{The expression for $\dddot{x}^\mu$ under MP condition} \label{app:mp}
Take the equations \eqref{eq:mpd} and \eqref{eq:pxdot} to express
\begin{align}
(m \dot{x}^\mu + \dot{x}_\gamma \dot{S}^{\gamma\mu})\dot\, = -\frac{1}{2} R^\mu_{\;\nu\kappa \lambda} \dot{x}^\nu S^{\kappa \lambda}\,.
\end{align}
Now use $S^{\mu\nu} \dot{x}_\nu = 0$ along with its time-derivatives and the fact that $\dot{x}_\gamma \ddot{x}^\gamma=0$, $\dot{x}_\gamma \dot{P}^\gamma=0$ to obtain \citep{pirani1956}
\begin{align}
m \ddot{x}^\mu - \dddot{x}_\gamma S^{\gamma \mu} = -\frac{1}{2} R^\mu_{\;\nu\kappa \lambda} \dot{x}^\nu S^{\kappa \lambda}\,. \label{eq:pir}
\end{align}
We now contract the expression above with $S_{\nu\mu}/S^2$ and partially re-express the result using the spin vector $s^\lambda$ to obtain 
\begin{align}
\dddot{x}^\kappa \left(\delta^\nu_\kappa + \dot{x}_\kappa \dot{x}^\nu - \frac{s_\kappa s^\nu}{S^2} \right) = \frac{m}{S^2} \ddot{x}^\mu S^\nu_{\;\mu} + \frac{1}{2 S^2} R_{\mu \lambda \kappa \gamma} \dot{x}^\lambda S^{\nu\mu} S^{\kappa \gamma} \,.
\end{align}
That is, we now have the expression for the jerk $\dddot{x}^\nu$ on the subspace orthogonal to $s^\lambda, \dot{x}^\kappa$. The projection of the jerk into velocity can be computed from the second derivative of four-velocity normalization as $\dddot{x}^\mu \dot{x}_\mu = - \ddot{x}^\mu \ddot{x}_\mu$. For the projection of the jerk into the spin vector, we use the Fermi-transport property $\dot{s}^\mu = - \dot{s}^\nu \dot{x}_\nu \dot{x}^\mu$ to express $\dot{s}^\nu \ddot{x}_\nu = 0$. This allows us rewrite the projection as
\begin{align}
 s^\mu\dddot{x}_\mu = \frac{\D}{\di \tau} (s^\mu \ddot{x}_\mu ) \,.
\end{align}
Now let us project Eq.~\eqref{eq:pir} into $s^\mu$ to obtain
\begin{align}
s^\mu \ddot{x}_\mu  = -\frac{1}{2m} R_{\mu \nu\kappa \lambda} s^\mu \dot{x}^\nu S^{\kappa \lambda}\,.
\end{align}
We now see that the time-derivative of $\ddot{x}^\mu s_\mu$ can be completely expressed by known functions of $x^\mu,\dot{x}^\nu,\ddot{x}^\kappa,s^\lambda$.

From that, it is now easy to compose the complete prescription for the jerk only in terms of the variables $\dot{x}^\mu,\ddot{x}^\lambda,s^\gamma$ as 

\begin{align}
\begin{split}
& \dddot{x}^\nu = \frac{1}{S^2}\left(m \ddot{x}_\mu - {\star R}_{\mu\lambda \kappa \gamma} \dot{x}^\lambda s^\kappa \dot{x}^\gamma  \right) \epsilon^{\nu\mu \sigma \tau}   \dot{x}_\sigma s_\tau  + \ddot{x}^\kappa \ddot{x}_\kappa \dot{x}^\nu \\
& + \frac{1}{m S^2} \left(  {\star R}_{\mu\lambda\kappa \gamma;\sigma} s^\mu \dot{x}^\lambda s^\kappa \dot{x}^\gamma \dot{x}^\sigma + 2{\star R}_{\mu\lambda\kappa \gamma} s^\mu  s^\kappa \dot{x}^{(\lambda} \ddot{x}^{\gamma)}   \right) s^\nu \,. 
\end{split}
\label{eq:trojak}
\end{align}



\section{Constraining the Khriplovich Hamiltonian} \label{app:cons}


\subsection{Constraint theory}

Let us first introduce some elements of Dirac-Bergmann constraint theory as presented, e.g., by \citet{dirac1966,hanson1976}. 

Let $\Phi^a = 0$ be a set of constraints on phase space we want to impose on the system, with $a$ some index labeling the constraints. Let us further assume that the matrix $C^{ab} \equiv \{\Phi^a, \Phi^b\}$ is non-degenerate and we can thus find an inverse matrix $C^{-1}_{ab}$. The goal is to find a Hamiltonian $H'$ which fulfills $\{\Phi^a,H'\} = \dot{\Phi}^a \cong 0$, where $\cong$ denotes an equality which is fulfilled under the condition that all the constraints $\Phi^a=0$ hold. Such a Hamiltonian can be obtained from the original one as
\begin{align}
H' = H - \{H,\Phi^a\}C^{-1}_{ab} \Phi^{b} \,.
\end{align}
In our particular case we will be imposing the constraints of the form $S^{\mu\nu} V_\nu = 0$. By counting the components of the constraint, we might be tempted to state that there are a total of 4 constraints imposed on the system. However, two components of the constraint are satisfied trivially due to the identities $S^{\mu\nu} V_\nu V_\mu = 0$ and $S^{\mu\nu} V_\nu {\star S_{\mu\kappa}} V^\kappa = 0$. As a consequence, the matrix $C^{\mu\lambda} \equiv \{S^{\mu\nu} V_\nu,S^{\lambda\kappa} V_\kappa \}$ will be degenerate on subspaces corresponding to these trivial constraints. However, it can be easily seen that if we find any pseudo-inverse $C^\dagger_{\mu\lambda}$, then the following Hamiltonian will conserve the non-trivial parts of the constraint and thus also the whole set $S^{\mu\nu} V_\nu = 0$ 
\begin{align}
H' = H - \{H,S^{\mu\nu} V_\nu\}C^{\dagger}_{\mu\lambda} S^{\lambda\kappa} V_\kappa \,.
\end{align}
The last note to this procedure is that in the following we never constrain the Poisson algebra; in other words, the Poisson brackets are always those given in \eqref{eq:poiss}. More details about this topic are discussed in the main text in Section \ref{sec:canon}.


\subsection{Obtaining the TD Hamiltonian} \label{subsec:tdcons}

The first constraint that we apply to the Hamiltonian (\ref{eq:kripham}) is $S^{\mu\nu} P_\nu = 0$. The constraint algebra yields
\begin{align}
&\{S^{\mu\nu} P_\nu, S^{\kappa\lambda} P_\lambda\} \cong -\tilde{\mathcal M}^2 S^{\mu \kappa}\,,\\
& \tilde{\mathcal M}^2 \equiv -g^{\mu\nu}P_\mu P_\nu + \frac{1}{4} R_{\mu\nu\kappa \lambda} S^{\mu\nu} S^{\kappa \lambda} \,.
\end{align}
The pseudo-inverse of $S^{\mu\kappa}$ on the constrained phase space is $- S_{\nu\mu}/S^2$ (cf. eq. (\ref{eq:proj})). The last bracket that needs to be evaluated is
\begin{align}
\{H_\mathrm{KS},S^{\kappa \lambda} P_\lambda\} \cong \frac{1}{2m} R_{\mu\nu\gamma \chi}S^{\kappa \mu} P^\nu S^{\gamma \chi} \,.
\end{align}
The constrained Hamiltonian then reads
\begin{align}
\begin{split}
H_\mathrm{TD} &= \frac{1}{2\mu} g^{\mu\nu} P_\mu P_\nu + \{H_\mathrm{KS}, S^{\kappa \lambda} P_\lambda\} \frac{1}{\tilde{\mathcal M}^2 S^2} S_{\mu\kappa} S^{\mu\nu} P_\nu 
\\
& = \frac{1}{2\mu} \left( g^{\mu\nu} + \frac{1}{\tilde{\mathcal M}^2}R^{\mu}_{\; \chi \xi \zeta}S^{\chi \nu} S^{\xi \zeta} \right) P_\mu P_\nu\,, \label{eq:Htdcons}
\end{split}
\end{align}
where we can apply $\cong$ equalities for expressions multiplied by the constraint $S^{\mu\nu} P_\nu$ without changing the resulting equations of motion. We have also chosen to change the notation $m \to \mu$ because as we will see, the meaning of the parameter $\mu$ will be different from the definition \eqref{eq:mdef}.  This Hamiltonian generates the equations of motion parametrized by some parameter $\lambda$ which does not need to be equal to proper time $\tau$. The equations of motion read

\begin{align}
& x'^\mu \cong \frac{1}{\mu} \left( g^{\mu\nu} + \frac{1}{2\tilde{\mathcal M}^2}R^{\nu}_{\; \chi \xi \zeta}S^{\chi \mu} S^{\xi \zeta} \right) P_\nu \,,\\
& P'^\mu \cong -\frac{1}{2} R^\mu_{\;\nu \kappa \lambda} x'^\nu S^{\kappa \lambda} \,,\\
& S'^{\mu \nu} \cong P^\mu x'^\nu - P^\nu x'^\nu \,,
\end{align}
where we denote the derivatives $\D/\mathrm{d} \lambda$ by primes. By comparing the equations above with the MPD equations of motion under the TD supplementary condition \eqref{eq:tdup} we see that the parameter $\lambda$ fulfills
\begin{align}
&\frac{\mathrm{d}\lambda}{\mathrm{d} \tau} = \frac{\mu m}{\mathcal{M}^2} \,,
\end{align}
where we substitute Eq.~\eqref{eq:mtd} for $m$. Another way to characterize the parametrization under the condition that $P^{\alpha}P_\alpha = -\mathcal{M}^2 = -\mu^2$ is that it holds that $P^\alpha x'_\alpha/\mathcal{M} = -1$. This is exactly the parametrization introduced by \citet{dixon1970} and vouched for by \citet{ehlers1977} \citep[see also][]{georgiosparam}. 
The Hamiltonian for world-lines parametrized by proper time is discussed in the main text in Subsection \ref{subsec:hamtd}. One should compare the above-given constraint procedure with the analogous constraint procedure in the vector-variable model of \citet{ramirez2015}.


\subsection{Other attempts}
We attempted to use the MP momentum-velocity relation \eqref{eq:mppu} and thus to apply the constraint $S^{\mu\nu}(\delta^\kappa_\nu + S^{\kappa \lambda} S_{\lambda \nu}/S^2) P^\nu =0$. The problem is, however, that once the spin tensor is degenerate, the identity $S^{\mu\nu}(\delta^\kappa_\nu + S^{\kappa \lambda} S_{\lambda \nu}/S^2) = 0$ holds automatically and has no time derivative under the Kriplovich Hamiltonian. In other words, the MP condition expressed in terms of momenta is satisfied by any degenerate spin tensor and it cannot be used in our constraint procedure.

The Corinaldesi-Papapetrou condition $S^{\mu\nu}\xi_\nu = 0$, where $\xi^\nu(x^\mu)$ is now some fixed vector field, can be applied as a constraint to yield the Hamiltonian
\begin{align}
H = \frac{1}{2m} g^{\mu \nu} P_\mu P_\nu + \frac{1}{{m \xi^2}} \xi_{\nu;\gamma} P^\gamma S^{\nu \kappa} \xi_\kappa \,.
\end{align}
Yielding the equations of motion
\begin{align}
& x''^\mu = -\frac{1}{2m} R^\mu_{\;\nu \kappa \lambda} x'^\nu S^{\kappa \lambda} - \frac{1}{\xi^2} \xi_{\nu;\gamma} x'^\gamma S^{\nu \kappa} \xi_{\kappa}^{\;\,;\mu} \,,\\
& S'^{\nu\kappa} = -\frac{1}{\xi^2} \xi_{\lambda;\gamma} x'^\gamma (S^{\lambda \nu} \xi^\kappa - S^{\kappa \lambda} \xi^\nu ) \,. 
\end{align}
Nevertheless, this set of equations are not the MPD equations under the Corinaldesi-Papapetrou condition. 


\section{Construction of canonical coordinates} \label{app:canon}
Consider the effective action for spinning bodies given in Ref. \citep{steinhoff2009}\revisiontwo{, written in terms of $p_\mu$ instead of $P_\mu$ using Eq.~\eqref{eq:pnonc}}:
\begin{align}
\mathcal{S} = \int \di \tau \left[ p_\mu \dot{x}^\mu + \frac{1}{2} S_{AB} \Omega^{AB} - H \right] \,, \label{eq:akce}
\end{align}
where $\Omega^{AB} \equiv \Lambda^A_{\;\;\hat{A}} \frac{\di \Lambda^{B \hat A}}{\di \tau}$ and $\Lambda^A_{\;\;\hat A}$ are the components of the ``body-fixed frame'' with respect to the background tetrad $e^A_{\mu}$. The body-fixed frame is defined by the property that the spin tensor is constant in it, $S^{\hat{A} \hat{B}} = \mathrm{const.}$, and $\Lambda^A_{\;\;\hat A}$ thus in fact carry the dynamical state of the spin tensor along with gauge degrees of freedom. We further assume here, unlike in Refs. \citep{steinhoff2009, steinhoff2015,vines2016}, that the Hamiltonian $H$ is only a function of the gauge-independent $p_\mu, x^\nu,S^{AB}$. It is then easy to show that the equations of motion following from $\delta \mathcal{S} = 0$, where $p_\mu, x^\nu, \Lambda^A_{\;\;\hat A}, S^{AB}$ are varied independently,
\revisiontwo{read
\begin{equation}
\dot{x}^\mu = \frac{\partial H}{\partial p_\mu}, \quad
\dot{p}_\mu = - \frac{\partial H}{\partial x^\mu}, \quad
\dot{S}^{AB} = 2 \Omega^{[A}{}_D S^{B]D}
  = 4 \frac{\partial H}{\partial S^{CD}} \eta^{C[A} S^{B]D} .
\end{equation}
This is equivalent to}
\begin{align}
\frac{\di f}{\di \tau} = \{f, H \}\,,
\end{align}
where $f$ is any function of $p_\mu, x^\nu,S^{AB}$ and the bracket is given as in Eq.~\eqref{eq:partcan}. \revisiontwo{Transforming to variables $x^\mu, P_\nu, S^{\mu\nu}$ via Eqs.~\eqref{eq:stetr} and \eqref{eq:pnonc} then leads to the Poisson brackets in Eqs.~\eqref{eq:poiss}.} In this sense, our Hamiltonian-based approach can be understood, up to the discarding of the $\Lambda^A_{\;\;\hat{A}}$ variables, as equivalent to the action-based approach of \citet{steinhoff2009,steinhoff2015,vines2016}. 

\revisiontwo{The action \eqref{eq:akce} with a given Hamiltonian $H$ should be understood as belonging to a fixed supplementary spin condition. However, based on this action, one can formulate the choice of supplementary condition as a gauge freedom, at least for the TD, CP/NW, and related supplementary conditions. This approach is used in Refs. \cite{Levi:2015msa,Steinhoff:2015ksa,Levi:2018nxp,vines2016}. The action \eqref{eq:akce} with the substitution of the KS and MP Hamiltonians \eqref{eq:kripham} and \eqref{eq:mpham} should also fit within this scheme but we leave a detailed investigation of this question for future work. 

Let us now return to the question of canonical coordinates.} We realize that if the term $S_{AB} \Omega^{AB}/2$ in the action can be transformed into the form $\sum_{i} \rho_{i}  \dot{\chi}^i$ with $\rho_i, \chi^i$ some dynamical variables, then $\rho_i, \chi^i$ are the desired pairs of canonically conjugate coordinates on the phase space. To do so, we mimic the approach presented in Ref. \citep{tessmer2013} and re-express
\begin{align}
\frac{1}{2}S_{AB} \Omega^{AB} = \frac{1}{2}S_{\hat{A} \hat{B}} \Lambda^{\;\;\hat A}_A \Lambda^{\;\;\hat B}_B \Omega^{AB} = -\frac{1}{2}S_{\hat{A} \hat{B}} \Lambda^{\;\;\hat A}_A \frac{\di \Lambda^{A\hat B}}{\di \tau}\,.
\end{align}
In other words, we are now looking at the dynamics of the spin tensor purely from the perspective of a Lorentz transformation $\Lambda^A_{\;\;\hat A}$ from the body-fixed frame into the referential tetrad. 

We now choose the spin tensor in the body-fixed frame to have one degenerate time-like direction and one non-degenerate space-like direction; conventionally $S_{\hat{1}\hat{2}} = - S_{\hat{2}\hat{1}} = S$ and other components zero. Note that this assumes that the spin tensor will eventually fulfill a supplementary spin conditions of the form $S^{\mu\nu}V_\nu=0$; non-degenerate spin tensors will thus not be possible to express in terms of the coordinates that we give in the following paragraphs.

To enable an intuitive discussion, let us further identify the legs $\Lambda^A_{\;\;\hat{1}},\Lambda^B_{\;\;\hat{2}},\Lambda^C_{\;\;\hat{3}}$ with the $x,y,z$-axes in Cartesian coordinates, and the $\Lambda^D_{\;\;\hat{0}}$ with the time axis. Then, by finding the dual of the spatial part of the spin tensor, we see that it is a vector of magnitude $S$ pointing purely in the $z$-direction. 

The spin tensor is invariant with respect to rotations around the $z$-axis, and with respect to boosts in the $z$ direction. Out of the total 6 parameters of a general Lorentz transform $\Lambda^A_{\;\;\hat A}$, 2 will be gauge degrees of freedom of the body-fixed tetrad. In order to not mix the gauge degrees of freedom and the true dynamical degrees of freedom, we parametrize the general Lorentz transform as
\begin{align}
\Lambda = R(\alpha,\vec{n}_z)B(v_z,\vec{n}_z)B(u,\vec{n}_\psi)R(-\theta, \vec{n}_\phi)\,,
\end{align}
where $R(\zeta,\vec{n})$ stands for a rotation by angle $\zeta$ around $\vec{n}$, and $B(v,\vec{n})$ a boost in the $\vec{n}$ direction. The numbers $\alpha,v_z,u,\psi,\theta,\phi$ are then generally time-dependent parameters of the transformation, and the vectors $\vec{n}_{\psi}, \vec{n}_{\phi}$ are given as
\begin{align}
\vec{n}_{\psi} = (\sin \psi,\cos \psi,0)\,,\\
\vec{n}_{\phi} = (\sin \phi,\cos \phi,0)\,.
\end{align}
When the dust settles, this transformation yields 
\begin{align}
&\frac{1}{2}S_{AB} \Omega^{AB} 
= -S \Lambda^{\;\hat 1}_A \frac{\di \Lambda^{A \hat{2}}}{\di \tau} 
\\ \nonumber &
= S\dot{\alpha} + S\frac{\cos\theta-1}{\sqrt{1 - u^2}}\dot{\phi} + S\left(\frac{1}{\sqrt{1 - u^2}}-1\right) \dot{\psi} \,.
\end{align}
The $S \dot{\alpha}$ term is a total time derivative and so it will not contribute to the equations of motion. From the other terms we see that we have two canonical momenta $A$ and $B$ conjugate to $\phi$ and $\psi$ respectively defined through the parameters of the Lorentz transformation as
\begin{align}
& A = S\frac{\cos\theta-1}{\sqrt{1 - u^2}}\,, \\
& B = S\left(\frac{1}{\sqrt{1 - u^2}}-1\right)\, \label{eq:BSfact}.
\end{align}
Expressions for these coordinates in terms of the components of the spin tensor are given in the main text in equation \eqref{eq:spcanon}. The expressions for the spin tensor components in terms of $A,B,\phi,\psi$ are then given in equation \eqref{eq:spparam}.

\subsection{Coordinate singularities and the special-planar Hamiltonian}

Imagine a particle moving along $x=0$ and $y=0$ in Cartesian coordinates in Euclidean space, and make the usual transform to spherical coordinates $r,\vartheta,\varphi$. In principle, the coordinate $\varphi = \arctan(x/y)$ is not defined, and we are at $\vartheta = 0$ or $\vartheta = \uppi$ depending on the sign of $z$. By a limiting procedure $x \to 0, y \to 0$, we are able to obtain any value between 0 and $2 \uppi$ for $\varphi$ at the pole. 

However, it is clear to us from the point of view of the more fundamental Cartesian coordinates that nothing is wrong, as the value of $\varphi$ is of no consequence for them at $\vartheta=0$. Similarly, $\dot{\varphi}$ is not defined at the pole, and by taking the azimuthal angular momentum along with $\vartheta$ to zero, we obtain any value for $\dot{\varphi}$ between $-\infty$ and $+\infty$; again, this is of no physical consequence and evolving $\varphi$ is redundant. 

The singularity at the pole of spatial spherical coordinates is similar to the singularity of the canonical coordinates for the spin tensor at $S^{A3} = 0$. By inspecting the transformation laws (\ref{eq:spcanon}) we see that the coordinate $\phi = \arctan (S^{31}/S^{23})$ is undefined and we are either at $A = 0$ or $A = -2(B + S)$ depending on the sign of $S^{12}$. 

In the case $A = 0$ ($S^{12} >0$), we see from the parametrization of the spin tensor (\ref{eq:spparam}) that the value of $\phi$ will in fact be of no consequence to the spin tensor. These conclusions can then be easily applied to an evolution that fulfills $S^{A3} = \rm const. = 0$ to reduce the number of variables we need to evolve.

In the case $A = -2(B+S)$ ($S^{12} <0$) the situations is somewhat more complicated. If we have an evolution that keeps $S^{A3} = \rm const. = 0$, we will also have $\dot{S}^{A3} = 0$. This, however, leads only to $\dot{A} = -2 \dot{B}$, and it is in fact the combination $2 \phi - \psi$ that uniquely parametrizes the spin tensor. For practical purposes, it then useful to define new canonical coordinates $D \equiv A/2 - B, E \equiv A/2 + B, \delta \equiv 2\phi - \psi, \epsilon \equiv 2 \phi + \psi$ so that $D,\delta$ and $E,\epsilon$ are conjugate respectively. The equation $\dot{S}^{A3} = 0$ with $S^{12} <0$ leads to $\dot{E} = 0$ and the redundance of the coordinate $\epsilon$.

For the special planar problem in Sec. \ref{sec:plan}, we chose $S^{12}>0$ for simplicity. A trick that can be eventually used to avoid the redefinitions of coordinates is simply to permute the definition of the tetrad elements $1 \leftrightarrow 2$, which will lead to a change of the physical meaning of the sign of $S^{12}$.

Another singularity is at $S^{A0} = 0$ which unambiguously leads to $B=0$ and $\psi$ undefined. Once again, we see in \eqref{eq:spparam} that the value of $\psi$ is inconsequential in that case. An interesting fact is that if we have an evolution such that $S^{A0} = \rm const. = 0$, then the coordinates $A, \phi$ reduce just to the canonical coordinates for the $\mathrm{SO}(3)$ Poisson algebra \citep[e.g.][]{lukes2014}.

\end{document}